\documentclass[sigconf,nonacm]{acmart}

\usepackage{placeins}
\usepackage{tabularx}
\usepackage{graphicx}
\usepackage{caption}
\usepackage{xstring}
\usepackage{etoolbox}
\usepackage{caption}
\usepackage{enumitem}
\usepackage{hyperref}
\usepackage{float}
\usepackage{subcaption}
\usepackage{xparse}
\usepackage{totpages}
\usepackage{multirow}
\usepackage{pgffor}
\usepackage{wasysym}
\usepackage{svg}
\usepackage{microtype}
\usepackage{cuted}

\usepackage{xspace}
\xspaceaddexceptions{]\}}

\newcommand{\df}[1][]{{DeckFlow}\ifx\relax#1\relax\else s\fi\xspace}
\newcommand{\cf}[1][]{{ChatFlow}\ifx\relax#1\relax\else s\fi\xspace}
\newcommand{\hand}[1][]{Hand\ifx\relax#1\relax\else s\fi\xspace}
\newcommand{\gc}[1][]{Goal~Card\ifx\relax#1\relax\else s\fi\xspace}
\newcommand{\card}[1][]{Card\ifx\relax#1\relax\else s\fi\xspace}
\newcommand{\ac}[1][]{Action~Card\ifx\relax#1\relax\else s\fi\xspace}
\newcommand{\tc}[1][]{Text~Card\ifx\relax#1\relax\else s\fi\xspace}
\newcommand{\ic}[1][]{Image~Card\ifx\relax#1\relax\else s\fi\xspace}
\newcommand{\cl}[1][]{Cluster\ifx\relax#1\relax\else s\fi\xspace}
\newcommand{\core}[1][]{AI Core\ifx\relax#1\relax\else s\fi\xspace}
\newcommand{\pname}{Chenxi\xspace}

\newcommand{\nparticipants}[2]{%
  n\textsubscript{%
    \ifnum#1=1%
      A%
    \else\ifnum#1=2%
      B%
    \fi\fi%
  }\kern.1em=\kern.1em#2\xspace%
}

\newcommand{\participantref}[2]{%
  P\textsubscript{%
    \ifnum#1=1%
      A%
    \else\ifnum#1=2%
      B%
    \fi\fi%
  }\kern.1em#2\xspace%
}

\newcommand{\paren}[1]{%
  \mbox{(#1)}\xspace
}

\usepackage{footnotehyper}
\makesavenoteenv{table}
\makesavenoteenv{tabular}

\author{Gregory Croisdale}
\affiliation{
  \institution{University of Michigan}
  \city{Ann Arbor}
  \country{USA}}
\email{gregtc@umich.edu}

\author{Emily Huang}
\affiliation{
  \institution{University of Michigan}
  \city{Ann Arbor}
  \country{USA}}
\email{emihuang@umich.edu}

\author{John Joon Young Chung}
\affiliation{
  \institution{Midjourney}
  \city{San Francisco}
  \country{USA}}
\email{jchung@midjourney.com}

\author{Anhong Guo}
\affiliation{
  \institution{University of Michigan}
  \city{Ann Arbor}
  \country{USA}}
\email{anhong@umich.edu}

\author{Xu Wang}
\affiliation{
  \institution{University of Michigan}
  \city{Ann Arbor}
  \country{USA}}
\email{xwanghci@umich.edu}

\author{Austin Z. Henley}
\affiliation{
  \institution{Carnegie Mellon University}
  \city{Pittsburgh}
  \country{USA}}
\email{azhenley@cmu.edu}

\author{Cyrus Omar}
\affiliation{
  \institution{University of Michigan}
  \city{Ann Arbor}
  \country{USA}}
\email{comar@umich.edu}

\renewcommand{\shortauthors}{Croisdale et al.}

\date{May 2025}

\begin{document}

\title{DeckFlow: Iterative Specification on a Multimodal Generative Canvas}

\begin{abstract}

Generative AI promises to allow people to create high-quality personalized media. Although powerful, we identify three fundamental design problems with existing tooling through a literature review.
We introduce a multimodal generative AI tool, DeckFlow, to address these problems. First, DeckFlow supports task decomposition by allowing users to maintain multiple interconnected subtasks on an infinite canvas populated by cards connected through visual dataflow affordances. Second, DeckFlow supports a specification decomposition workflow where an initial goal is iteratively decomposed into smaller parts and combined using feature labels and clusters. 
Finally, DeckFlow supports generative space exploration by generating multiple prompt and output variations, presented in a grid, that can feed back recursively into the next design iteration. We evaluate DeckFlow for text-to-image generation against a state-of-practice conversational AI baseline for image generation tasks. We then add audio generation and investigate user behaviors in a more open-ended creative setting with text, image, and audio outputs.

\end{abstract}

\begin{teaserfigure}
\vspace{-.5pc}
    \includegraphics[width=\textwidth]{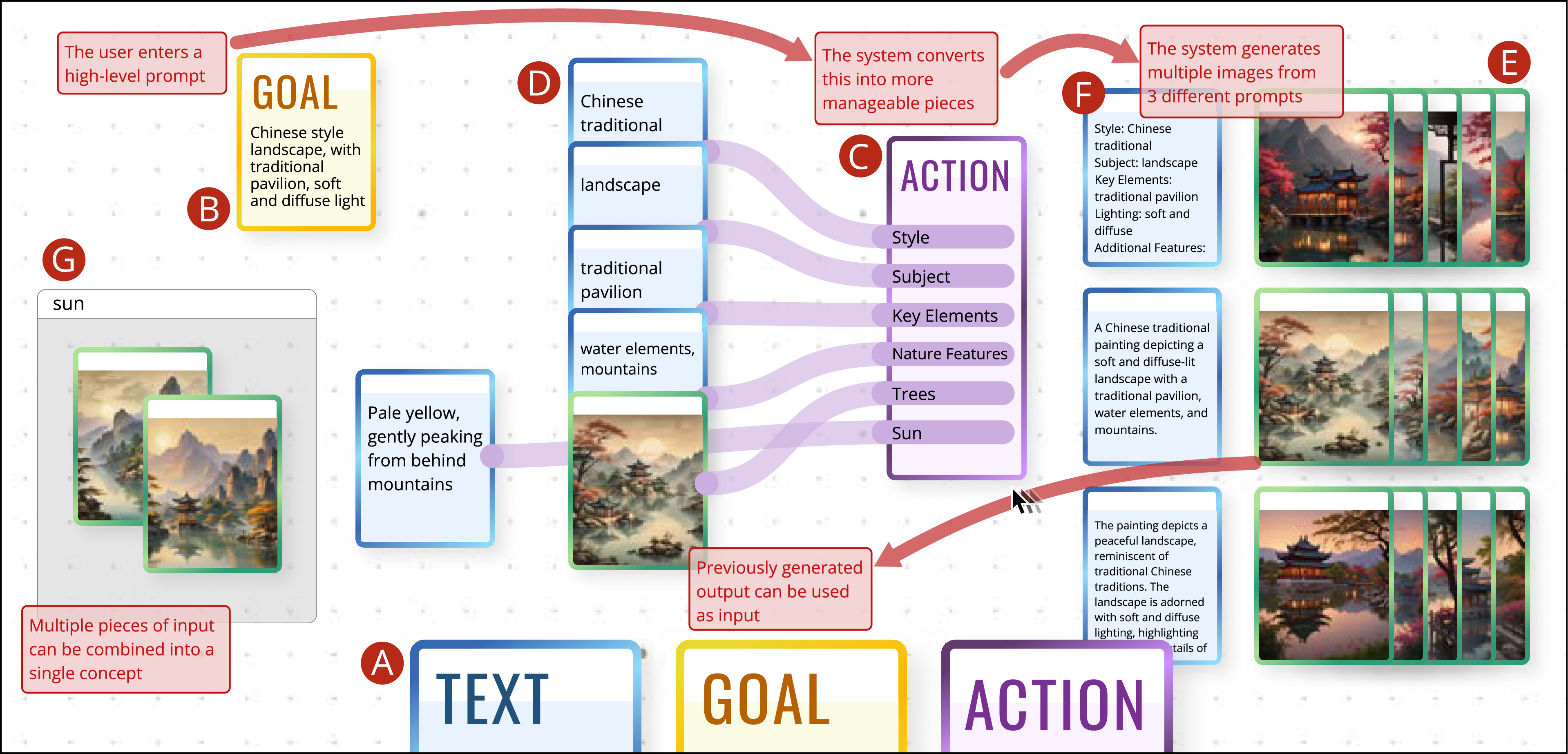}
\vspace{-1.5pc}
\caption{
\df is an infinite canvas for creating multimodal content.
    In this case, detailed in Section \ref{sec:motivating_example}, the user drags a \gc~(b) from the \hand~(a), which generates an \ac~(c) connected to several \tc[s]~(d) representing the decomposed specification. The \ac spawns multiple \tc[s] containing the constructed prompts (f), and images are generated using them (e) so the user can explore the generative space. In a subsequent iteration of the task, the user moves some of them into a \cl~(g), and uses one as input to the \ac.
}
\label{fig:teaser}
\Description{Illustration showing DeckFlow's cards on a large canvas. Various inputs and outputs of text and images.}
\end{teaserfigure}



\maketitle

\section{Introduction}\label{sec:introduction}

With a short text prompt, someone with minimal prior knowledge can generate a clever Shakespearean poem about a jaunt on a sunny day in ChatGPT, an image of a beautiful oil painting of a flowing wheat field at sunset in the style of Van Gogh in Midjourney\footnote{Midjourney. \href{https://www.midjourney.com/home}{https://www.midjourney.com/home}}, or a catchy pop-punk song pontificating about global warming in Suno\footnote{Suno. \href{https://suno.com/}{https://suno.com/}}. How can simple prompts lead to such complex and compelling output? Generative AI models rely on statistical patterns derived from massive amounts of publicly available training data. The model makes assumptions about the user's intent based on the most commonly observed patterns in the training data. Training data exhibits the same biases as online content in general \cite{zhou2024biasgenerativeai}, however, so this sort of one-shot prompting is limited in its ability to creatively support diverse users and niche use cases. 

In situations where the model's output is unsatisfying, the user may want to iteratively refine their specification and try again. However, common generative AI tools make it difficult to exert fine-grained control over what is generated and are limited in their support for iteration. Users often resort to randomly tweaking the prompt and re-running the model until the generated output is acceptable, or they simply give up. 

This experience has led to an explosion in research on user interface affordances that provide more fine-grained control over generative AI and better support for creative iteration. 
We start in Section \ref{sec:background} with a survey of prior work on generative AI tools, including tools for generating media of various modalities, such as images, audio, text, and code. 

Often, users want to generate artifacts that consist of sub-parts, and which would require multiple sub-tasks that separately engage a generative AI model. For example, when designing a scene in a fantasy narrative, the user might want to generate a fantasy setting, a creature, and a soundtrack separately before combining the results to form the final artifact of interest. Many generative AI tools are limited in their support for multiple parallel but connected sub-tasks (the \textbf{task decomposition problem}). 

Within each sub-task, users might want to combine more than one prompt, constraint, or example to express their intent, but existing tools are limited in their ability to combine multiple specifications focusing on different aspects of a single task (the \textbf{specification decomposition problem}).
For example, users may want to provide a natural language prompt describing a creature's basic form---say ``a fantasy dragon''---while providing an image as an example of the body they want the creature to have, a group of other images to describe the creature's eyes more specifically, and another group of images as examples of the general art style they would like. 

After providing an initial specification of their intent, users ``roll the dice'' 
by letting the generative AI model generate output, often multiple times. Generative AI is stochastic in nature, so each output differs slightly, forming a space of possible outputs for a given specification. Users may be interested in exploring this space, but many existing tools only present one output, forcing the user to sequentially ask the model to generate multiple outputs (the \textbf{generative space exploration problem}). 

To address these problems, this paper introduces DeckFlow. 
DeckFlow is a \textbf{multimodal generative AI tool} designed to support a variety of creative activities. Our focus in this paper was on working with text, images, and audio, both as input to and as output from the tool. \hyperref[fig:teaser]{The teaser image} shows a simplified example of a user workflow.

To address the \textbf{task decomposition problem}, all creative activity occurs on an infinite canvas consisting of a collection of cards connected using visual dataflow affordances. Multiple sub-tasks, or different iterations of a task, can be performed in parallel on the canvas by separating the tasks spatially, or interacting with each other through connections.

To address the \textbf{specification decomposition problem}, 
DeckFlow supports a specification workflow where an initial Goal Card, typically consisting of a text prompt, is decomposed into an Action Card, consisting of several textual labels, which serve as ``ports'' in the dataflow diagram. Text, image, and audio cards can be connected to these ports. The system initially decomposes the Goal Card into a more granular collection of Text Cards to initialize the Action Card.

To address the \textbf{generative space exploration problem}, users can click a button on an Action Card to generate three groups of three outputs from the underlying generative AI model. Each output appears directly on the infinite canvas next to the corresponding Action Card upon request. The user can freely delete, group, or rearrange these outputs to explore the design space. The user can also freely repurpose the output from one iteration of content generation as input to one or more other tasks, including future iterations of the same task.

\begin{enumerate}
    \item Section \ref{sec:background} provides an analysis of existing literature and tools related to AI-assisted content generation.

    \item Section \ref{sec:Deckflow} introduces DeckFlow, a multimodal generative AI tool that contributes novel solutions to each of the central problems as just described.

    \item Sections \ref{sec:evaluation}-\ref{sec:results} present empirical evaluations through two comprehensive studies:
    \begin{enumerate}
        \item A comparative study contrasts DeckFlow with a conversational baseline (ChatFlow) in open-ended and closed-ended image generation tasks, demonstrating the ability to afford distinct decomposition styles, rich interactivity in the decomposition of specifications, and similar outcomes with replication tasks, but significant improvements in creative tasks.
        \item A multimodal behavioral study extends DeckFlow with audio generation and input, finding that users decompose open-ended creative tasks which involve multiple modalities in similar, structured ways, demonstrating the tendency to rely upon text in specification decomposition, and the distinct challenges associated with multimodal exploring generative spaces.
    \end{enumerate}
\end{enumerate}

In addition to our validation of \df as a whole, this paper contributes generalizable knowledge in the form of 
(1) our decomposition of the design space of generative AI tools around the three central problems that organize every section of this paper, 
(2) a set individual affordances in \df, validated by our study, that could be implemented in other visual generative AI tools, e.g. Action cards and our lightweight generative space exploration affordances, and 
(3) insights about how humans engage in task and specification decomposition and generative space exploration in multimodal content generation tasks.

\section{Background}\label{sec:background}

\subsection{Task Decomposition}\label{bg:decomposition}

Producing contemporary creative artifacts, like feature films, video games, and mixed‑media installations, requires coordinating distinct but interdependent tasks to produce and combine artifacts of various modalities (video, audio, 3D geometry, code). Contemporary creative tools therefore have developed affordances to support \emph{task decomposition}, i.e. decomposing larger tasks into smaller sub-tasks.

For example, the open‑source \emph{Blender} spans the 3D stack (modeling, texturing, animation, lighting, etc.), while isolating individual tasks so, for example, lighting tweaks do not require considering textures~\cite{BlenderFoundation}.
Infinite canvas tools such as \emph{Figma} extend this principle spatially, allowing teams to cluster related frames while retaining a bird’s‑eye view of the whole project~\cite{Figma}. \emph{Code Bubbles} applies the same idea to code on an infinite canvas, improving developer understanding and reducing navigation time~\cite{bragdon2010code}. Ethnographic studies of developer whiteboards suggest that such spatial arrangements decrease working memory load and externalize spatial mental models~\cite{whiteboard}. 

Generative AI tools have also started to explore the problem of task decomposition. Conversational interfaces, like ChatGPT, enable task decomposition through the ability to create multiple conversations and conversation groups. The language model itself can further break down a larger task into a sequence of smaller sub-tasks, e.g. by being directed to use Chain-Of-Thought reasoning~\cite{schulhoff2024promptreportsystematicsurvey}. 

A number of other generative AI tools have explored infinite canvases for task decomposition. Some tools retain a chat-like prompting strategy, but arrange outputs from different tasks visually on an infinite canvas. For example, \emph{Promptify} lays out each batch of image generation results on a zoomable canvas. These images arise from a textual prompt entered into a text box that the user modifies over time \cite{Promptify2023}.
Other systems retain the inputs themselves on the canvas, but not as individual entities, including \emph{ComfyUI}~\cite{comfyui}, \emph{ChainForge}~\cite{arawjo2023chainforge}, \emph{Sensecape} (a text-only tool)~\cite{sensescape}, and \emph{tldraw computer}~\cite{tldrawComputer}. Of these, \emph{Sensecape} and \emph{tldraw computer} combine both approaches, with both inputs and outputs appearing on the canvas and with the ability to repurpose prior outputs as inputs for a subsequent task.
DeckFlow also takes this approach to task decomposition.

\subsection{Specification Decomposition}\label{bg:spec-decomp}

Within each sub‑task, users often need to specify several distinct aspects of the creative artifact, like its style, palette, tone, or rhythm. Conventional creative tools provide a variety of affordances specialized to each of these. However, many contemporary generative AI tools require expressing every aspect of the artifact using a natural language prompt. Practitioners therefore improvise, e.g. by including bullet‑point lists or pasted reference images, but in some domains, this can limit their ability to specify their intent precisely, e.g. with regard to a particular aspect of an image while leaving others unchanged.

Generative AI tools have contributed affordances that help address this \emph{specification decomposition problem}. \emph{ChainForge} can construct a prompt from a template string. These fields can be independently swept or frozen, enabling controlled A/B testing across a single dimension~\cite{arawjo2023chainforge}. \emph{CreativeConnect} lets users specify discrete keywords connected to specified regions of a sketch~\cite{choi2023creativeconnect}. 
\emph{CueFlik} frames specification decomposition as interactive concept learning: users label positive and negative image examples, and the system learns a weighted combination of visual features that can be re‑applied across queries~\cite{CueFlik}. 
\emph{PromptPaint} interpolates continuously between multiple prompts during the diffusion process, exposing a weighted blend rather than a concatenated string~\cite{chung2023promptpaint}.
\emph{PromptCharm} provides a mixed‑initiative loop: an RL‑based agent suggests refined prompts, while users can tweak token‑level attention or in-paint masked regions, exposing prompt, attention, and pixel masks~\cite{wang2024promptcharm}. 

In the domain of strictly textual tools, \emph{Sensecape} allows users to decompose prompts into individual parts, arranged spatially, and compare variants in parallel, merging the parts they like~\cite{sensescape}. \emph{Luminate} asks writers to tag sentences along qualitative dimensions (e.g., formality, concreteness) and then recombines those dimensions to generate tailored drafts~\cite{suh2024luminate}. 

The regex synthesis tool \emph{Regae} shows that letting users iteratively add examples or constraints to a live candidate set, rather than authoring a monolithic spec, reduces cognitive load and speeds convergence~\cite{regaeGlassman}. Empirical work on \emph{dimensional reasoning} similarly reports that separating axes of variation aids sense‑making, even though it introduces a modest interface learning barrier~\cite{dimensional_reasoning}.

\df contributes novel affordances to support specification decomposition in multimodal generative AI workflows. The user initially specifies a high-level Goal Card, which the system first decomposes into an Action Card that has labeled ports for each of these features as well as initial input cards connected to each port. Users can then modify or create new variations of each card as they perform their task. 

\subsection{Generative Space Exploration}\label{bg:gen-explore}

Generative AI models are stochastic and can generate a wide variety of outputs for a given task specification.
Long before modern generative AI models, visualization researchers argued that creative work benefits from design galleries: curated arrays of parameter variations that reveal structure in high-dimensional spaces~\cite{Marks1997DesignGalleries}.  Graphic-layout tools such as DesignScape revived the idea for automatic poster composition, presenting users with multiple exemplar layouts and allowing them to steer by favoring particular variants~\cite{ODonovan2015DesignScape}. These gallery-based approaches exemplify a shift from producing a single ``best'' artifact to navigating a solution space. Modern generative models make that space substantially larger.

Rather than picking just one output from this space, many generative AI tools provide affordances for \emph{generative space exploration}. \emph{Sensecape} treats every generated response from part of a decomposed prompt as a movable card; users can request additional responses, duplicate or branch cards, build hierarchical concept maps, and thus form a multilevel mental model of the generative space~\cite{sensescape}.  \emph{Promptify} offers a lighter abstraction: each successive revision of a single prompt appears on a zoomable canvas, preserving visual history and encouraging lateral comparison, though genuine branching still requires manually copying the prompt~\cite{Promptify2023}.  \emph{ComfyUI}, aimed at experts, provides a direct manipulation approach which exposes seeds, schedulers, and CFG scales as node parameters so designers can sweep numeric ranges and cache intermediate latents, effectively turning low-level controls into gallery axes~\cite{comfyui}. \emph{Dreamsheets} gives a similar, but more accessible interface, adding image generation directly into a spreadsheet application~\cite{dreamsheets}.

\df turns exploration into a \emph{multimodal branching graph}. When triggered, each Action Card spawns three output branches, each based on a minor prompt variants and each itself containing three possible outputs. Any image, text, or audio card can be dragged back as a slot value for that action card, allowing iterative generative space exploration.

\section{DeckFlow}
\label{sec:Deckflow}

\newcommand{\subfig}[1]{\hyperref[fig:teaser]{(#1)}}

To address the design problems just identified, we designed and implemented \df to support task decomposition, specification decomposition, and enable exploration.

\subsection{Motivating Example}
\label{sec:motivating_example}
To begin, we will walk through an example \df usage scenario.
\pname has just moved into her new apartment and wants to decorate the dining room. She knows she wants a landscape picture showcasing the scenery of her hometown, in a style similar to a Chinese ink painting, but beyond that, is overwhelmed with possibilities and struggles to begin.

\pname starts by dragging from the \hand \subfig{a}, creating a \gc \subfig{b} in which she writes ``Chinese style landscape, with traditional pavilion, soft and diffuse light.'' The \gc then creates an \ac \subfig{c} with labels connected to discrete \tc[s] \subfig{d}, extracted from her high-level prompt: \texttt{Style}: ``Chinese traditional'', \texttt{Subject}: ``landscape'', \texttt{Key Elements}: ``traditional pavilion'', \texttt{Lighting}: ``soft and diffuse'', and \texttt{Natural Features}, but because these features weren't specified in the original prompt, this connection is empty. As a result, \pname thinks of natural features she is particularly interested in: ``water elements, mountains.'' Satisfied with these settings, \pname asks the \ac to generate some images from her specifications.

The \ac begins by creating three different prompts, using the labeled inputs as guidance. \pname is now presented with three rows of \ic[s] \subfig{e}, prompted using different \tc[s] \subfig{f}. The first row's \tc is created by simply concatenating the inputs together: ``Style: Chinese traditional, Subject: landscape...'', resulting in images which closely match her input. The second row's \tc is created by calling an LLM with those labeled inputs, creating a more coherent version of the prompt: ``A Chinese traditional painting depicting a serene landscape with a traditional pavilion...'' The third row uses these inputs in a small local LLM which has been optimized to generated creative and aesthetically appealing image prompts, yielding a less-precise, but more interesting row of \ic[s]: ``In the heart of a serene Chinese courtyard, a traditional Chinese painting unfolds, featuring a serene landscape...''

In a couple of the \ic[s] from the second row, \pname likes the way the sun is peaking through the mountains in the background, but isn't quite sure how to describe it. She moves these \ic[s] to a different region of the canvas and forms a \cl \subfig{g} around them, indicating to it that she wants to understand how to describe the phrase `sun' from these \ic[s]. After clicking the `interpret' button, the \cl generates a \tc illustrating how to create a prompt which captures this essence: ``Pale yellow, gently peaking from behind mountains.''

Satisfied with this description, \pname decides to modify the existing \ac to include this information. She begins by moving the old \ic[s] to a region above the \ac so that she can refer to them later. \pname adds the input `sun' to the \ac, and connects the \tc from the \cl to it. After regenerating, she realizes that the images are missing the pink cherry trees she liked from her previous set of images. Rather than using the \cl this time, she decides to add a new input, `trees', and connects an \ic to it.

With these changes, \pname decides to re-generate a new batch using the \ac. These next three rows of images provide different interpretations of her input. \pname finds that the second row matches her expectations better, finding one in particular that she likes the most.

\subsection{Interface Overview}

\begin{figure}[h]
    \centering
    \includegraphics[width=0.45\textwidth]{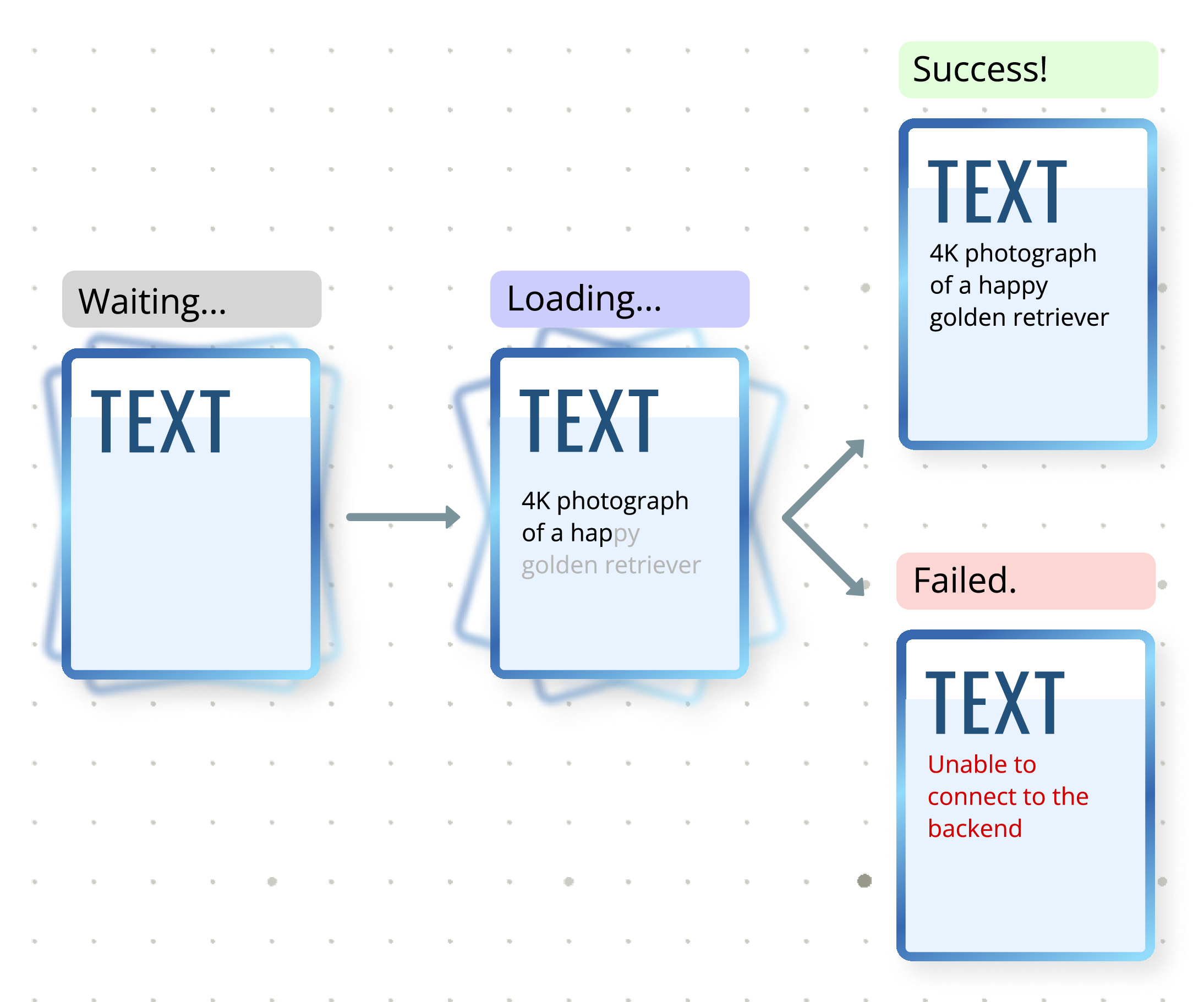}
    \caption{The lifecycle of a generated card. Micro-interactions help communicate the system status to the user.}
    \label{fig:card_states}
\end{figure}

In order to enhance engagement and familiarity within the system, UI elements and interactions were designed around a card-game aesthetic. To make the dynamic state of the system easily glanceable, cards have subtle animations and a message bubble (diagrammed in Figure \ref{fig:card_states}), inspired by the card game Cultist Simulator\footnote{Cultist Simulator. \href{https://weatherfactory.biz/cultist-simulator/}{https://weatherfactory.biz/cultist-simulator/}}.

\subsubsection{Task Decomposition on an Infinite Canvas}
\label{df:canvas}

Cards in \df are organized on an Infinite Canvas, inheriting several interface mechanics familiar to users of programs like Miro, Figma, or Powerpoint. Users can zoom or pan the view using a trackpad or mouse. Users can select cards by clicking each individually, or by dragging a rectangular region. These entities can then be dragged, duplicated, or removed. To create a connection between entities, users can drag an output socket into an input socket, snapping when close. Eligible entities can be placed into a \cl by pressing the ``cluster'' button on a selection. Selections can also be copied and pasted into the interface, or as serialized elements in other applications.
The infinite canvas is intended to support flexible Task Decomposition. 

\subsubsection{\card[s] as an Interactive Task Atom}
\begin{figure}[h]
    \centering
    \includegraphics[width=0.45\textwidth]{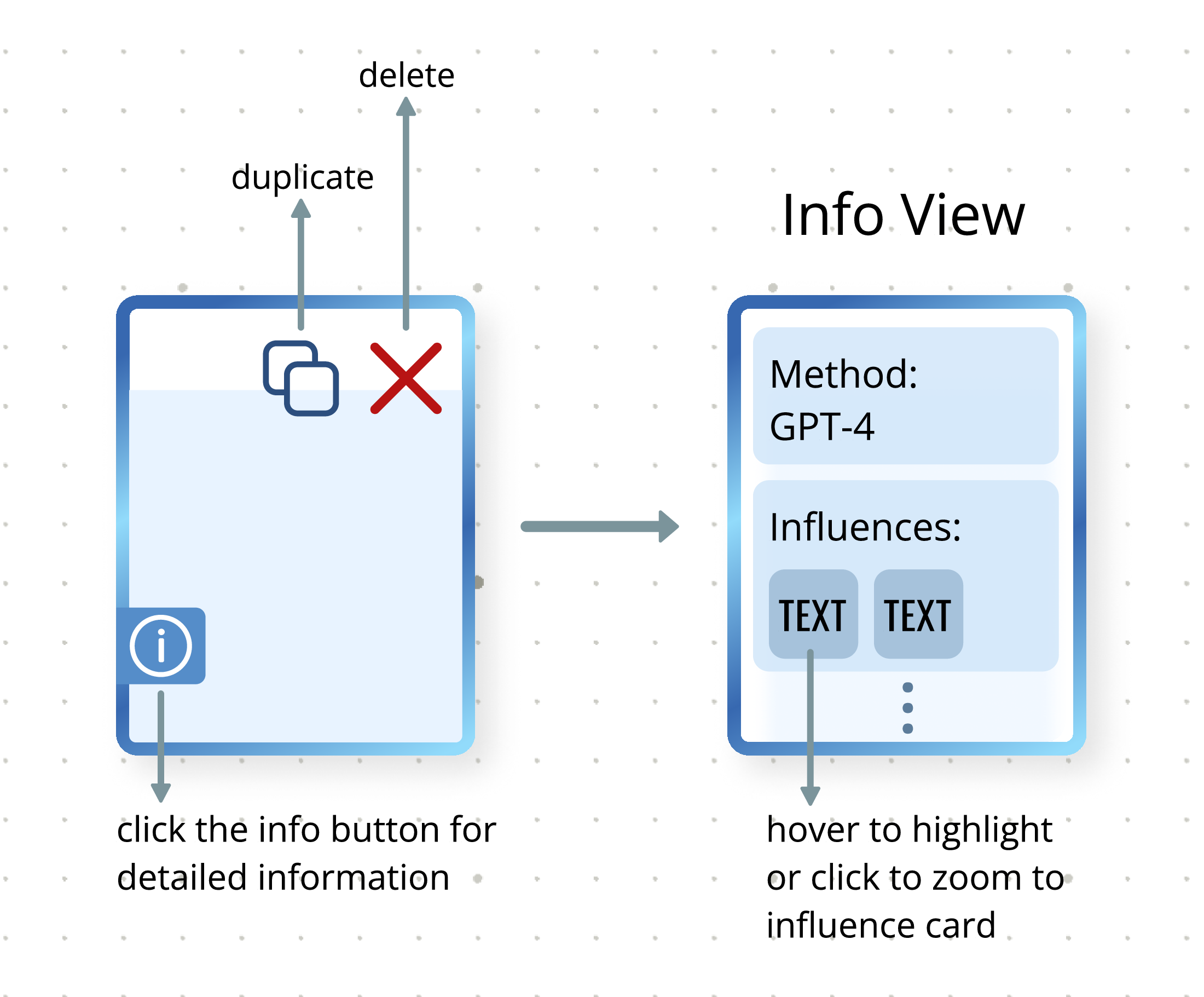}
    \caption{Cards have an Info Button which reveals information about how it was created, a Duplicate button and a Delete button.}
    \label{fig:card}
\end{figure}

To support system visibility, a \card can have one of the following states: `waiting', `loading', `error', or `success', as seen in Figure \ref{fig:card_states}. When in the `waiting' state, a \card will slowly shake, indicating that it is waiting for the backend to begin working on it. Once the backend has begun computation, it will shift to the `loading' state, shaking faster, and revealing a bubble to show what computation is occurring. Upon entering the `error' or `success' state, a small `jump' animation will play, and a bubble will appear explaining the transition. To promote recall and resumption, \card[s] also have a toggleable ``Info View'', as seen in Figure \ref{fig:card}, which provides information regarding its history on the interface, such as a clickable reference to the \card[(s)] which influenced its generation, and the method and prompt which was used to generate it.

\begin{figure*}[h]
 \centering
 \includegraphics[width=\textwidth]{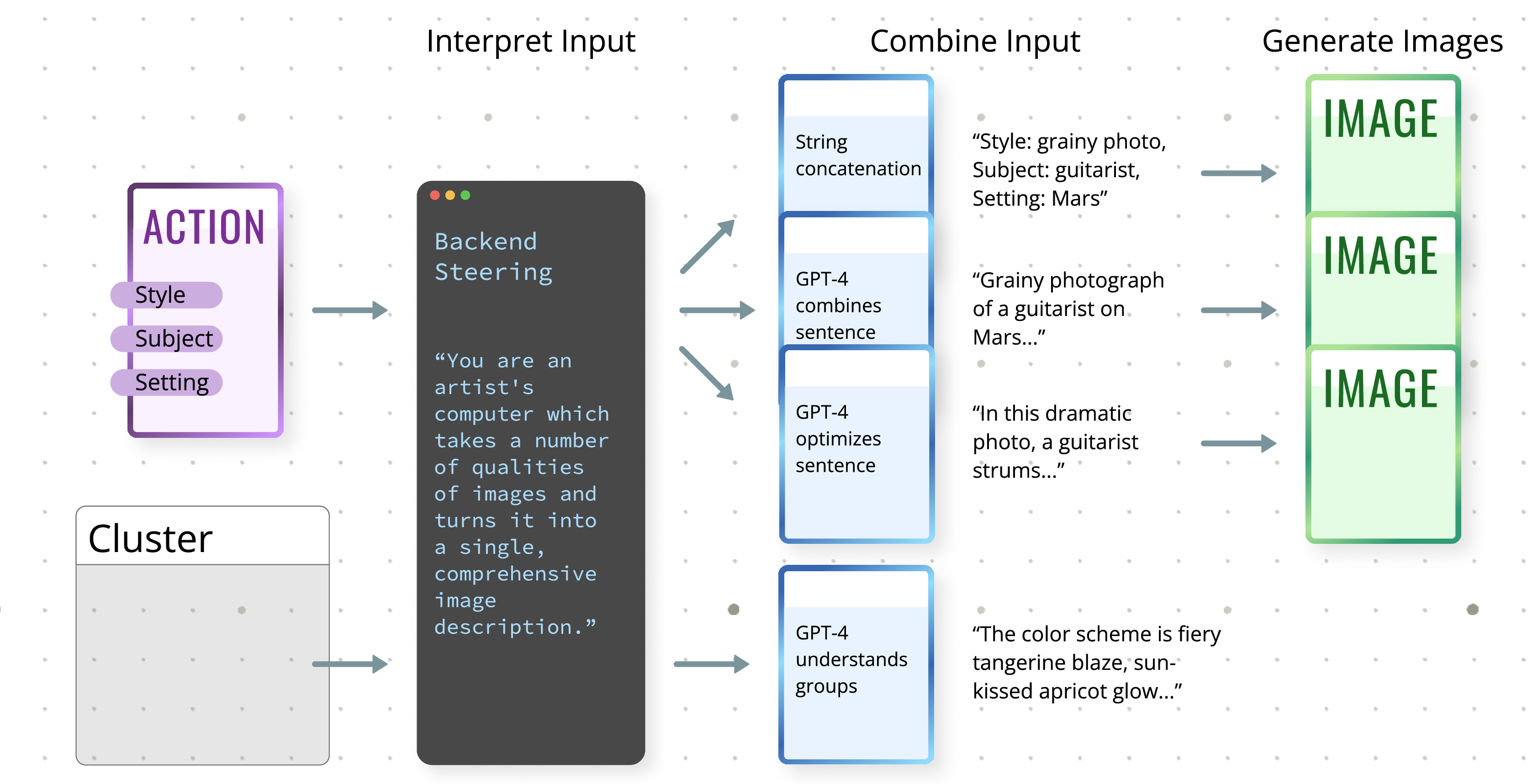}
 \caption{The \ac and \cl send similar requests to the \core, requesting input interpretation, combination, and for the \ac, subsequent Image Generation.}
 \label{fig:compute_chart}
\end{figure*}

At the bottom of the interface, there is a `Hand'  of cards~\subfig{a} which the user can drag from to create new elements. Additionally, users can drag files or paste data into the interface, and \df will convert them into a supported \card element.

\subsubsection{Data Cards}
In \df, all data is represented by a data element, either uploaded by the user or generated by the system. These elements can serve as input to any functional element which accepts a connection, with implicit type conversion performed using the user annotation, introducing multimodal input to all functional elements. There are three types of data elements, the \ic, the \tc, and the Audio Card, containing Image, Text, or Audio data respectively. A \tc can be edited by the user at any time, but the \ic and Audio Card is immutable once created.

\subsubsection{Function Cards}
Function elements allow the user to interact with data elements and provide input. There are three types of function elements: the \ac, the \gc, and the \cl.

\subsubsection{\ac~\subfig{c} for Specification Decomposition}
\label{df:actioncard}
Action cards have a set of sockets labeled with text, which we refer to as ``labels'' in this paper. These text labels can be used flexibly to support specification decomposition. For example, if an \ic containing a landscape image is connected to an \ac with the label `trees', then the generative AI model may interpret this as only the trees of the landscape should influence the generated \ic[s]~\subfig{h}. To programmers, this may be reminiscent of creating a function prototype, writing input variable names which are indicative of how they should be used to provide cognitive scaffolding.

When triggered, the \ac combines the content of all connected \card[s], creating three \tc[s]~\subfig{f}. At this point, if the \ac's target modality is not text, it uses these text cards to serve as the prompt for 3 cards of the target modality using \df's \core~\subfig{e}. 

\subsubsection{\gc~\subfig{b} for Collaborative Specification Decomposition}
\label{df:goalcard}
During pilot studies, users expressed frustration in creating \ac[s], unsure of where to get started. In response, we created the \gc. The \gc accepts goal text from the user using a text box, such as the teaser's \textit{``Chinese style landscape, with traditional pavilion, soft and diffuse light.''} Upon being triggered, it utilizes \df's \core to generate an \ac~\subfig{c} with text labels to serve as scaffolds, extracted from the provided goal text, already connected to a \tc matching if defined in the goal text~\subfig{d}, or disconnected to indicate to the user that it needs more information.

\subsubsection{\cl~\subfig{g} for Example-Based Specification}
\label{df:cluster}
Like the \ac, a \cl can accept arbitrary number of cards of any data type. The \cl, however, contains its inputs, moving them as it moves. Rather than accepting a text label for individual input, the \cl has space for only one optional text label, representing the idea of the entire cluster. When triggered, it uses \df's \core to generate a \tc which represents the shared features of each of the inputs, steered by the text label. Early prototypes of the \cl had an output socket which connected to an \ac directly, but pilot studies indicated that this lacked necessary transparency, as users wanted to make adjustments to the output, such as extracting a few keywords, before connecting to an \ac.

\begin{table*}[t]
  \centering
  \caption{Tasks utilized during the Comparative Study.}
  \renewcommand{\arraystretch}{1.1}
  \begin{tabular}{@{}p{3cm}p{4.5cm}p{4.5cm}@{}}
    \toprule
    \textbf{Task} & \textbf{A} & \textbf{B} \\ \midrule
    \textbf{Close‑Ended}\\recreate a given image as closely as possible &
    \raisebox{-\totalheight}{\includegraphics[width=4.3cm]{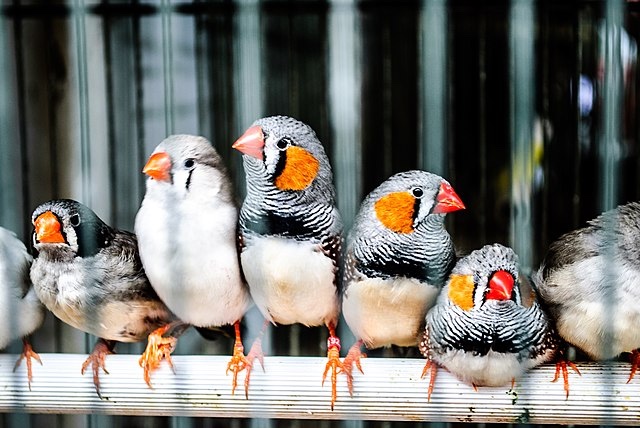}} &
    \raisebox{-\totalheight}{\includegraphics[width=4.1cm]{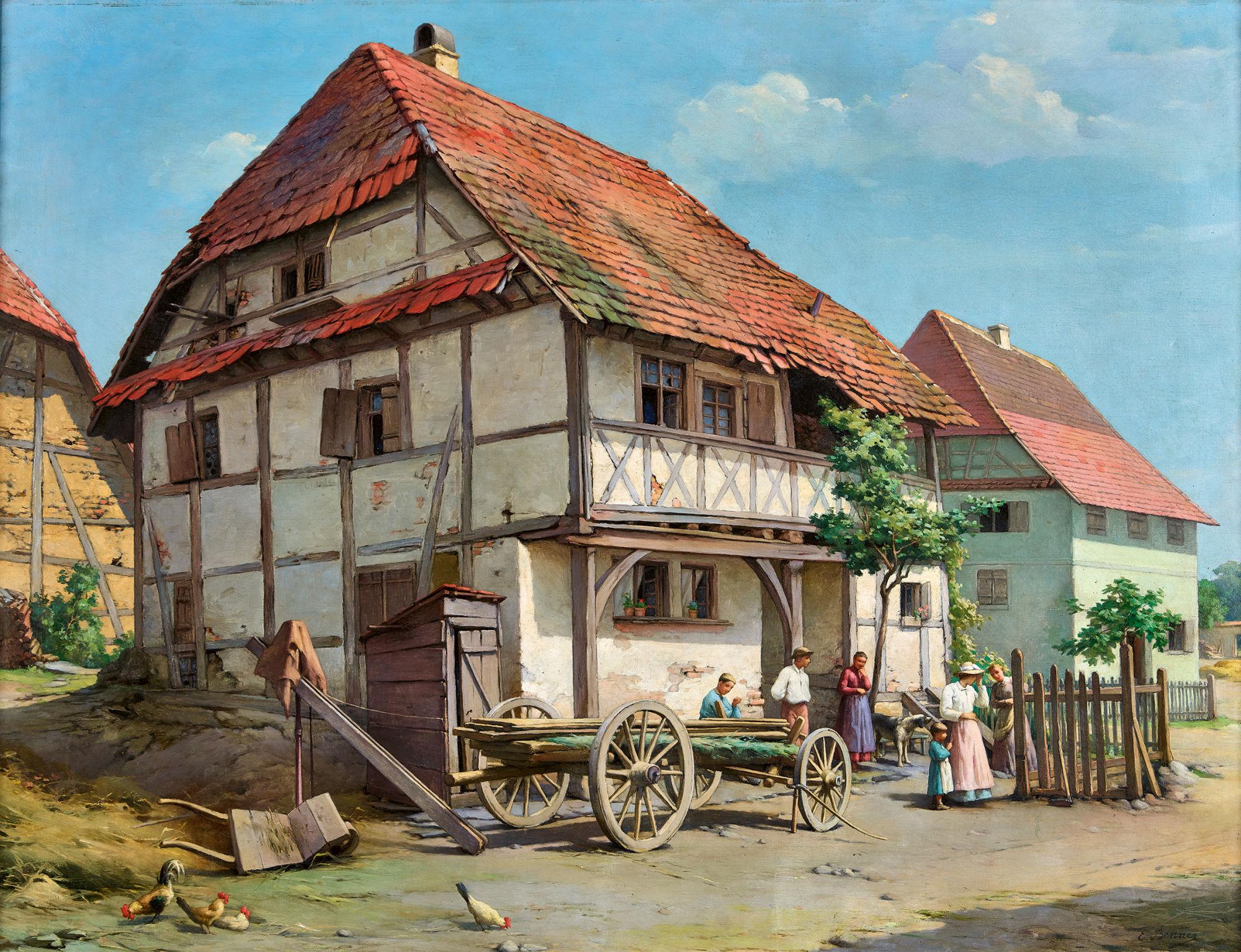}} \\ \addlinespace
    \textbf{Open‑Ended}\\create an image that best satisfies a text prompt &
    Create a picture you might like to hang up in your dining room &
    Create a picture of a place you might like to live \\ \bottomrule
  \end{tabular}
  \label{tab:comparative_tasks}
\end{table*}

\subsection{Implementation}
\df is an open-source application, which will be released publicly under the MIT license. The \df system consists of a HTML5 frontend, NodeJS backend, and hot-swappable workers in Python.

\subsubsection{Frontend}
\df's frontend was programmed in React, Typescript, and NodeJS. Early versions utilize tldraw\footnote{tldraw. \href{https://github.com/tldraw/tldraw}{https://github.com/tldraw/tldraw}} for the Infinite Canvas interactions. This was replaced with a custom Infinite Canvas implementation in order to overcome core system issues relating to circular state dependencies and a non-permissive license. All interface elements and their animations, such as the \card[s] and \cl[s], were written in React, Typescript, and Sass.

\subsubsection{Backend}
The backend, written in NodeJS, manages a database of shared data, a list of clients (i.e., instances of the \df interface), a list of workers which can perform computations at the request of the user, and a WebSocket server for them to communicate with each other. Upon receiving a request from the frontend, the backend chooses an available worker to process that request, prioritizing ones which have the required models already loaded.

\subsubsection{Workers}
To communicate with the backend, we created a Python library to perform work and update client elements in \df, designed to support future extensions and modalities. Each worker node utilizes this library to accept certain jobs from the backend, such as ``Generate Image'', ``Interpret Data'', or ``Generate Text'', illustrated in Figure \ref{fig:compute_chart}. These can be specialized according to available hardware and dynamic user needs. Worker nodes communicate data updates to the frontend through a WebSocket, while providing the frontend with real-time status updates.

\subsubsection{\core}
\label{df:core}
The \core handles most of the logic and computation in \df. For example, as shown in Figure \ref{fig:compute_chart}, it takes input from \ac[s], \cl[s] and other components, and interprets them multi-modally, combines the interpreted results in various ways for prompting diverse image generation, and finally prompts the Image Generation model to generate images.
To determine the settings for the \core, we performed a series of preliminary testing on a range of VLMs which support the image modality, system prompts, and one-shot examples, and eventually opted to use GPT-4-Vision-Preview\footnote{GPT-4-Vision-Preview. \href{https://platform.openai.com/docs/models/gpt-4-and-gpt-4-turbo}{https://platform.openai.com/docs/models/gpt-4-and-gpt-4-turbo}} for its performance and availability. As zero-shot acoustic audio understanding is not as available or robust as image understanding, the prompts which generated audio are used as input, but this can be easily replaced as audio AI develops.

\subsubsection{Image and Audio Generation}
To balance responsiveness and quality, Stable Diffusion XL Lightning\footnote{SD-XL Lightning. \href{https://huggingface.co/ByteDance/SDXL-Lightning}{https://huggingface.co/ByteDance/SDXL-Lightning}} was selected for Image Generation. To broadly improve prompt quality and divergence in the third row of the \ac's generated \ic[s], the 77M parameter LLM SuperPrompt\footnote{SuperPrompt. \href{https://brianfitzgerald.xyz/prompt-augmentation/}{https://brianfitzgerald.xyz/prompt-augmentation/}} was selected. For audio generation, the Stable Audio model \cite{evans2024stableaudioopen} was selected, as it is one of the few openly accessible audio generation models currently available; while Stable Audio excels at tasks like sound effects, it is less successful in creating legible outputs such as music or spoken word.

\section{Evaluation Method}\label{sec:evaluation}

\begin{table*}[ht]
 \centering
 \caption{Tasks utilized during the Multimodal Behavioral Study.}
 \begin{tabular}{| c | p{4cm} | p{4cm} | p{4cm} |}
 \hline
 \textbf{Task Name} & \textbf{Creature} & \textbf{Chess Club} & \textbf{Children's Book} \\ \hline
 \textbf{Media} &
 \raisebox{-\totalheight}{\includegraphics[width=4cm]{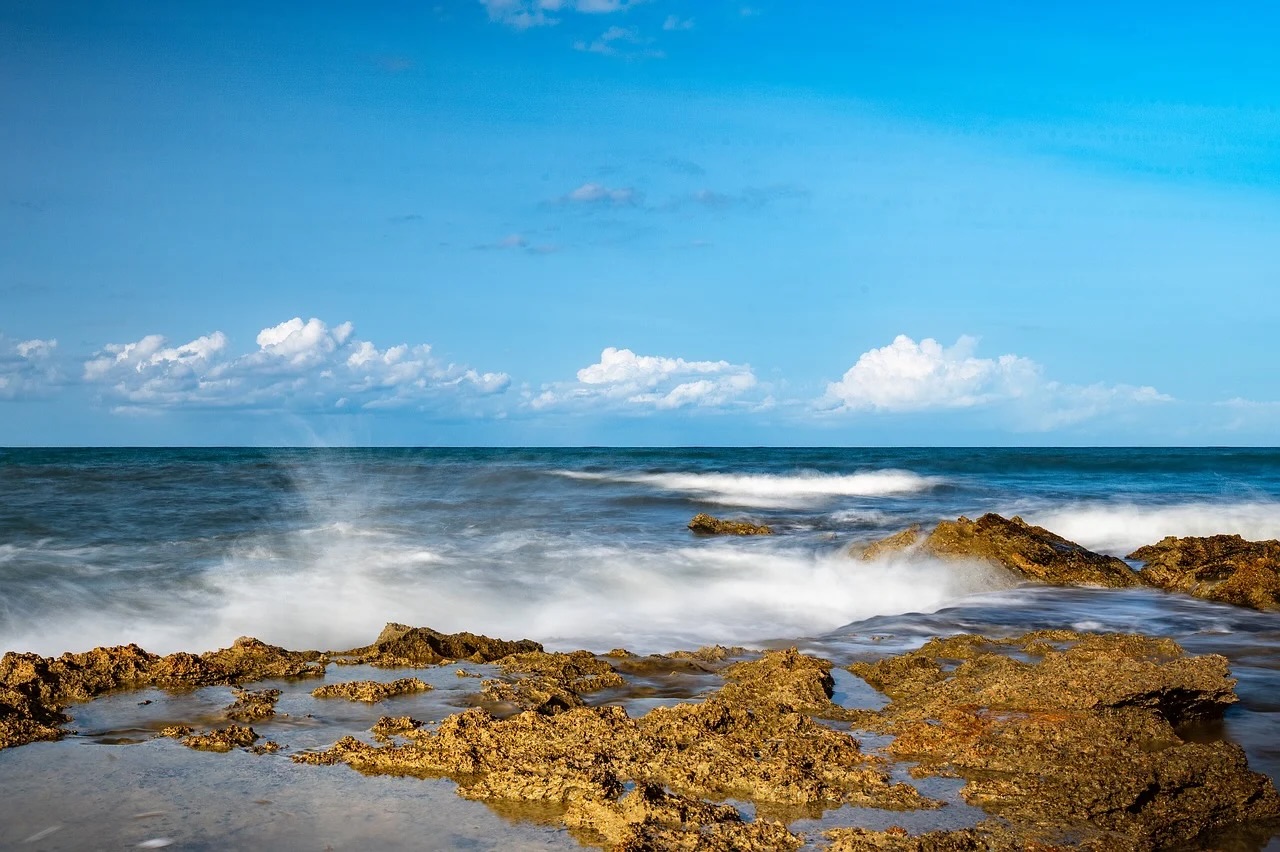}}  &
    {Keywords:\newline
    chess, fun, social, sun \newline
    Time: 3:00pm, June 4th, 2024 \newline
    Location: Green Park, 123 Main St
    } &
    Evening\_Melodrama.mp3\footnote{\href{https://incompetech.com/music/royalty-free/mp3-royaltyfree/Evening\%20Melodrama.mp3}{`Evening Melodrama'' Kevin MacLeod (incompetech.com). Licensed under Creative Commons: By Attribution 4.0 License.}}
    \\
 \hline
    \textbf{Instructions} &
        Imagine a creature has the following qualities:\newline
        It is considered "cute",\newline
        It has 6 legs,\newline
        and lives in the environment pictured above. &
        Create components of a media campaign to advertise this event. &
        You are writing a short children's story which might have this soundtrack. A page should include some image and some text.
 \\
 \hline
    \textbf{Output} &
    \begin{enumerate}
        \item An encyclopedia entry which describes the characteristics of the creature
        \item An image of the creature in its habitat
        \item An audio clip of the creature
    \end{enumerate} &
    \begin{enumerate}
        \item An image to advertise the event
        \item Text that you might include on a poster to encourage people to join, including the time and location
        \item A sound bite that you might include on an TikTok or Instagram Short for the event
    \end{enumerate} &
    \begin{enumerate}
        \item A page which describes the setting of your story
        \item A page which describes some minor conflict in the story
        \item A page which resolves that conflict in the story
        \item Some sound effect for one of the pages
    \end{enumerate}\\
 \hline
 \end{tabular}
 \label{tab:multimodal_tasks}
\end{table*}

To evaluate the effectiveness of DeckFlow, understand patterns of user behavior, and extract generalizable insights relevant to the designers of other generative tools targeting end-users, we conducted two studies. The first study compares DeckFlow with ChatFlow, a ChatGPT-like interface for text and image generation that uses the same backend generative AI models (the \textbf{comparative study}). This is followed by a more in-depth study of user behavior using a version of DeckFlow modified to also support audio (the \textbf{multimodal behavioral study}). Each session was scheduled to be 2 hours long, and participants were compensated with a \$30 USD gift card.

\subsection{Comparative Study}\label{sec:comparative_study}
When choosing a baseline, we considered a number of related tools. Many modern image generation interfaces described in the literature like Promptify~\cite{Promptify2023}, WorldSmith~\cite{dang2023worldsmith}, and GenQuery~\cite{son2023genquery} did not have publicly accessible code during the selection time. Others which offer promising interaction methods like Luminate~\cite{suh2024luminate} and ChainForge~\cite{arawjo2023chainforge} did not support image generation at the time of the study. The popular image generation interfaces Automatic1111~\cite{automatic} and ComfyUI~\cite{comfyui} are one-shot text-to-image generation systems, with little support for task or specification decomposition. Therefore, we chose the most powerful state-of-practice baseline: a conversational interface based on the widely used ChatGPT tool. To avoid comparing against an artificially weak strawman, we resolve some artificial limitations of ChatGPT in an interface called \cf, which allows users to generate arbitrarily many images at once, edit prior messages in the chat history without removing existing messages, and re-generate any output as needed. Additionally, ChatFlow uses the same image generation model, Stable Diffusion XL Lightning, as DeckFlow.

\subsubsection{Research Questions}
This study sought to understand how the novel interface, \df, impacted text to image generative tasks.

\noindent \textbf{RQ1:} How do users approach the Task Decomposition problem in both interfaces?\\
\textbf{RQ2:} How do users approach the Specification Decomposition problem in both interfaces?\\
\textbf{RQ3:} How do users approach the Generative Space Exploration problem in both interfaces?

\subsubsection{Procedure}

We conducted a within-subjects user study consisting of open and closed-ended text to image generation tasks with 16 students recruited from a Computer Science and Engineering email list: 8 male, 7 female, and 1 non-binary, all of whom were between 18--25 years of age. 

First, users were given a brief demographic survey, followed by an introduction interview. After this, users were given a brief tutorial in each interface. Then, users were given two think-aloud tasks, shown in Table \ref{tab:comparative_tasks}, for each tool with a flexible time limit of 10 minutes each (2 tasks $\times$ 2 tools $\times$ 10 minutes = 40 minutes). The tasks and the order of tools used were counterbalanced to avoid ordering effects.

\begin{enumerate}
\item A closed-ended task, where participants were asked to recreate a given image as closely as possible; these are designed to study situations in which a user's design space is narrowly defined.
\item An open-ended task, requiring participants to create an image that best satisfies a given text prompt; these are designed to study more exploratory, divergent design processes.
\end{enumerate}
This dual-task study design incorporates common techniques in evaluating image generation and modification tools, approximating realistic image tasks \cite{evirgen2022ganzilla, wang2024promptcharm}.

We collected data through various means, including basic usage metrics (e.g., number of images and text inputs used), user ratings of different features, self-reported success in ChatFlow vs. DeckFlow, screen and voice recordings, and interviews at the beginning, after each task, and at the conclusion of each study.

To qualitatively analyze this data, we watched each study several times, extracting utterances and notable uses of the tool, populating an affinity diagram. We then separated these findings into themes, including counter-examples, and crafted interpretations.

\subsection{Multimodal Behavioral Study}\label{sec:multimodal_study}
After analyzing the findings from the first study, we sought to further interrogate these findings and the extent to which they also apply to text and audio generative targets. We also took this opportunity to perform a design iteration to improve small aspects of the user interface where we observed users consistently struggled:
\begin{itemize}
    \item Audio was added as input and output
    \item \gc[s] were unified with \tc[s]
    \item Clusters were directly usable as input rather than merely generating a \tc
    \item \tc[s] could be dynamically sized to make them more visually manageable
\end{itemize}

We conducted a user study involving 7 individual participants, also recruited from a Computer Science and Engineering email list, of which 3 were male, 4 were female, all of whom were between 18--25 years of age. Users performed 2 tasks from a pool of 3, shown in Table \ref{tab:comparative_tasks}, in which they generated a text, image, and audio artifact for each.

\subsubsection{Research Questions}

This study was designed to more deeply understand the results from the Comparative Study, so we again evaluate RQs 1--3, evaluating only \df, and additionally explore the following question:

\noindent \textbf{RQ4:} How are different modalities treated as input and output?

\subsubsection{Procedure}

In this study, after a brief demographic survey followed by a introduction interview, users were given a brief tutorial of Deckflow. Then, users were given two think-aloud tasks, shown in Table \ref{tab:multimodal_tasks}, counterbalanced to include 2 of the tasks from a pool of 3. These tasks were administered with a flexible time limit of 15 minutes each. Some participants took substantially longer in these tasks, meaning they could only complete one task during the study period. 

After each task, state was reverted to the time in which a user used each of the primary input mechanisms: Text, Image, Audio, Cluster, as well as Action Card generation. If a user had not used one of the input mechanisms naturally, they were asked to perform another generation. During this retrospective think-aloud, users were asked questions relating to their expectations and their perception of the output.

At the conclusion of the study, we ran a modified version of the Creativity Support Index survey \cite{creativity_support_index} for each major feature: text/image/audio as input, cluster usage, and action card generation. We sought to understand user perception of these features of \df in order to perform a precise analysis.

We collected more targeted data in this study, including a detailed log indicating each action a user takes, along with screen and voice recordings. To classify the types of labels that were used by participants, we constructed a codebook with examples from the previous study.

\FloatBarrier

\section{Results}
\label{sec:results}

We detail the results of both user studies, comparing \df and the baseline ChatFlow using system logs, interview results, recordings of participant use, and survey results, in an effort to answer our research questions. To refer to participants, we use the format \participantref{1}{2}, where the subscript indicates the study (A for \hyperref[sec:comparative_study]{Comparative Study}, B for \hyperref[sec:multimodal_study]{Multimodal Study}), and the number is a unique identifier for that participant. To refer to participants, we use the format \nparticipants{2}{3}.

\subsection{RQ1: How do users approach the Task Decomposition problem in the interfaces?}\label{finding:task_decomposition}
One of the biggest differences between \df and other Generative AI interfaces, such as a conversational interface or a traditional dataflow programming environment, is the ability for users to organize their input and output in whatever way they desire, not constrained by data-flows, linear chat, or modal type conversions.

\subsubsection{Users developed distinct styles of use in \df}\label{finding:distinct}

\begin{figure*}[t]
 \centering
 \begin{subfigure}{0.3\textwidth}
 \includegraphics[width=\textwidth]{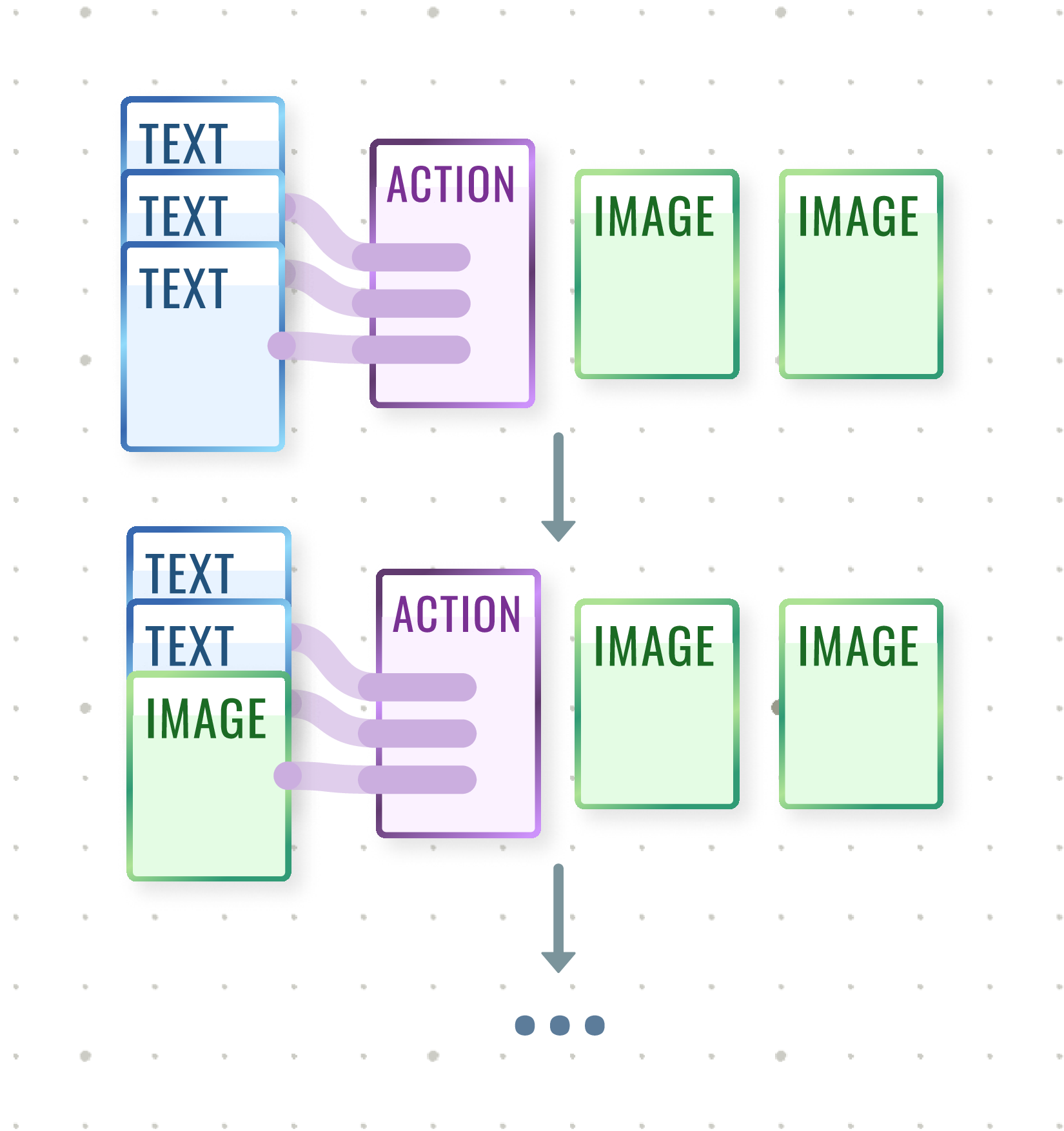}
 \caption{Top-Down Sequential Method \paren{\nparticipants{1}{9}}}
 \label{fig:org_methods:sequential}
 \end{subfigure}
 \begin{subfigure}{0.3\textwidth}
 \includegraphics[width=\textwidth]{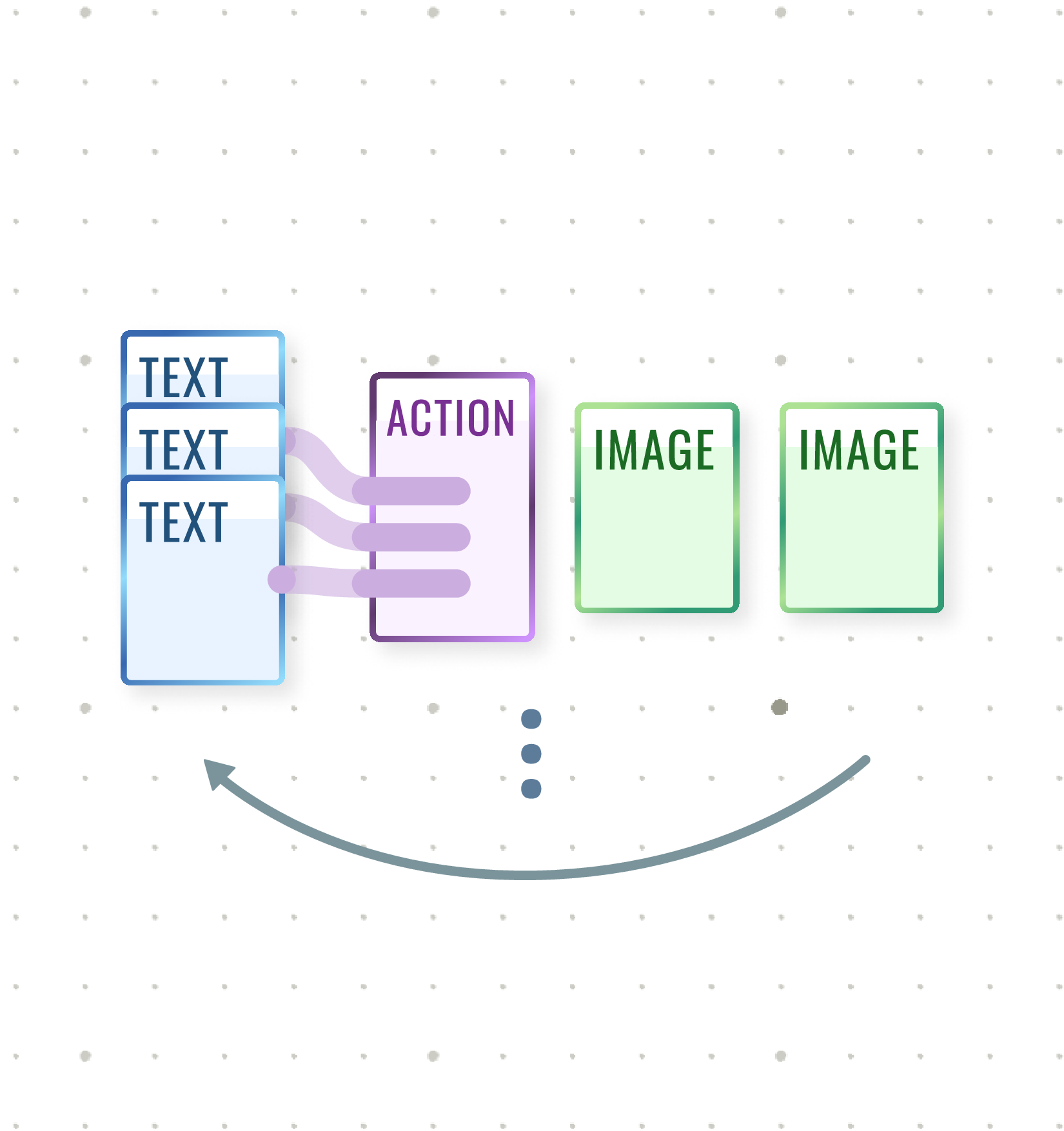}
 \caption{One-Card Iteration Method \paren{\nparticipants{1}{4}}}
 \label{fig:org_methods:onecard}
 \end{subfigure}
 \begin{subfigure}{0.3\textwidth}
 \includegraphics[width=\textwidth]{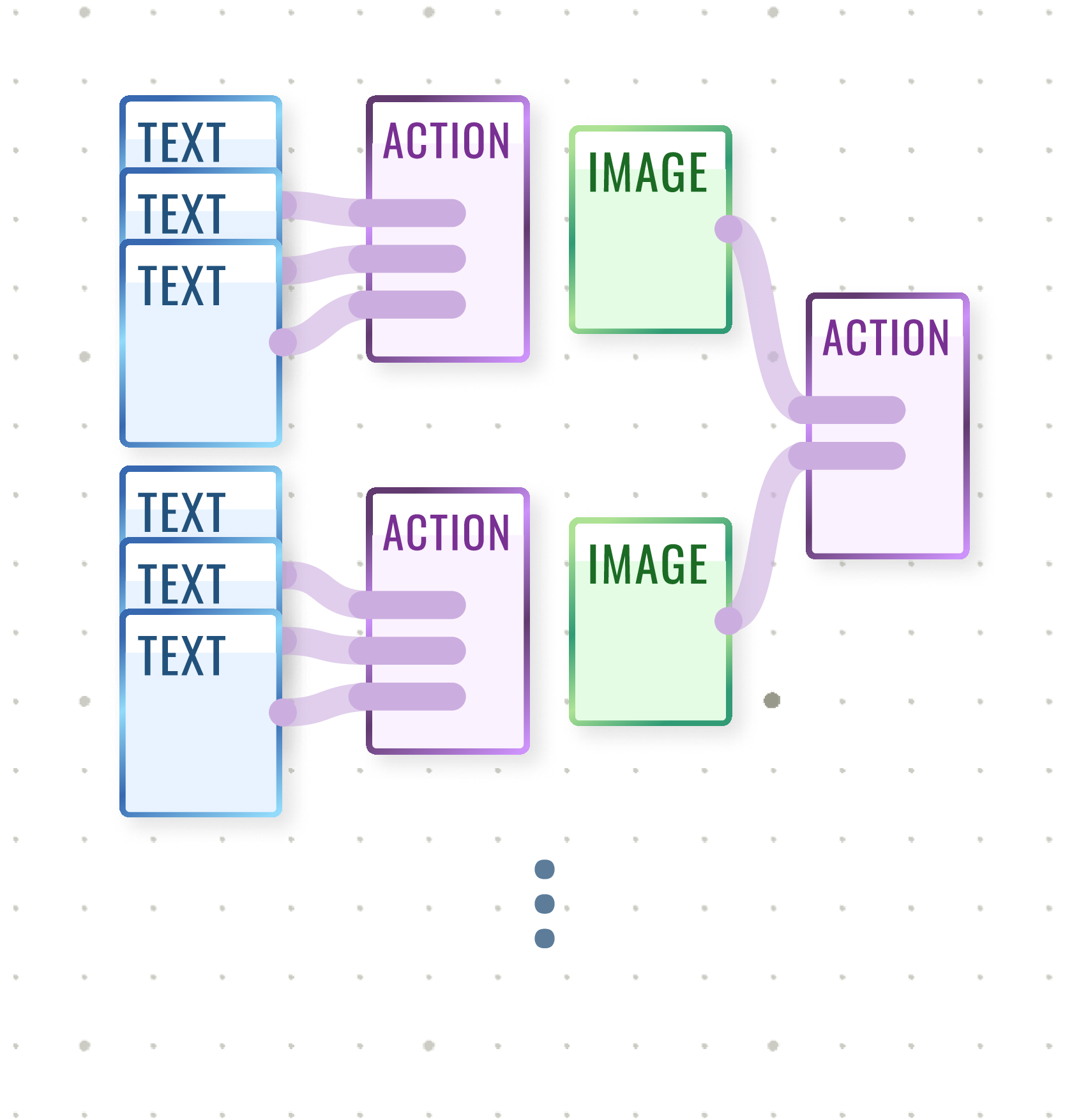}
 \caption{Divide-and-Conquer Method \paren{\nparticipants{1}{3}}}
 \label{fig:org_methods:divide}
 \end{subfigure}
 \caption{Different workflows in \df observed in the Comparative Study \paren{\nparticipants{1}{16}}, as discussed in \ref{finding:distinct}.}
 \vspace*{-1em}
 \label{fig:org_methods}
\end{figure*}

Upon detailed analysis of usage logs and study recordings of the Comparative Study, three spatial patterns of use emerged:

\begin{enumerate}
    \item \textit{Top-Down Sequential} \paren{\nparticipants{1}{9}}: Users such as \participantref{1}{6} used the interface linearly, in a manner comparable to the baseline, creating new \ac[s] connected to previous output and desired modifications in \tc[s] to iteratively improve upon their output. ``It was nice to copy things and have them right next to the old input, change a few things, and then go back up and see your old work'' \paren{\participantref{1}{6}}.
    \item \textit{One-Card Iteration} \paren{\nparticipants{1}{4}}: Users like \participantref{1}{5} further emphasized pure iteration in their approaches, using only one or two \ac[s] for their generations. ``I don't like moving things around, if I have to move hold down for an extended period of time'' \paren{\participantref{1}{5}}.
    \item \textit{Divide-and-Conquer} \paren{\nparticipants{1}{3}}: As exemplified by \participantref{1}{4}, this method consists of specializing different image features in different areas of the board, taking successful image or text prompts to a ``main'' area of the board upon reaching satisfactory prompt adherence. ``It's like a storyboard. I'm developing the architecture style here, and I'm gonna finally plug it into the main \ac'' \paren{\participantref{1}{4}}.
\end{enumerate}

Due to the increased number of outputs required in the Multimodality study, these trends were not apparent, but modal patterns emerged, explored in \ref{rq:modalities}.

\subsubsection{Clusters as a structural tool}\label{finding:cluster_structure}

\begin{figure*}[t]
 \includegraphics[width=\textwidth]{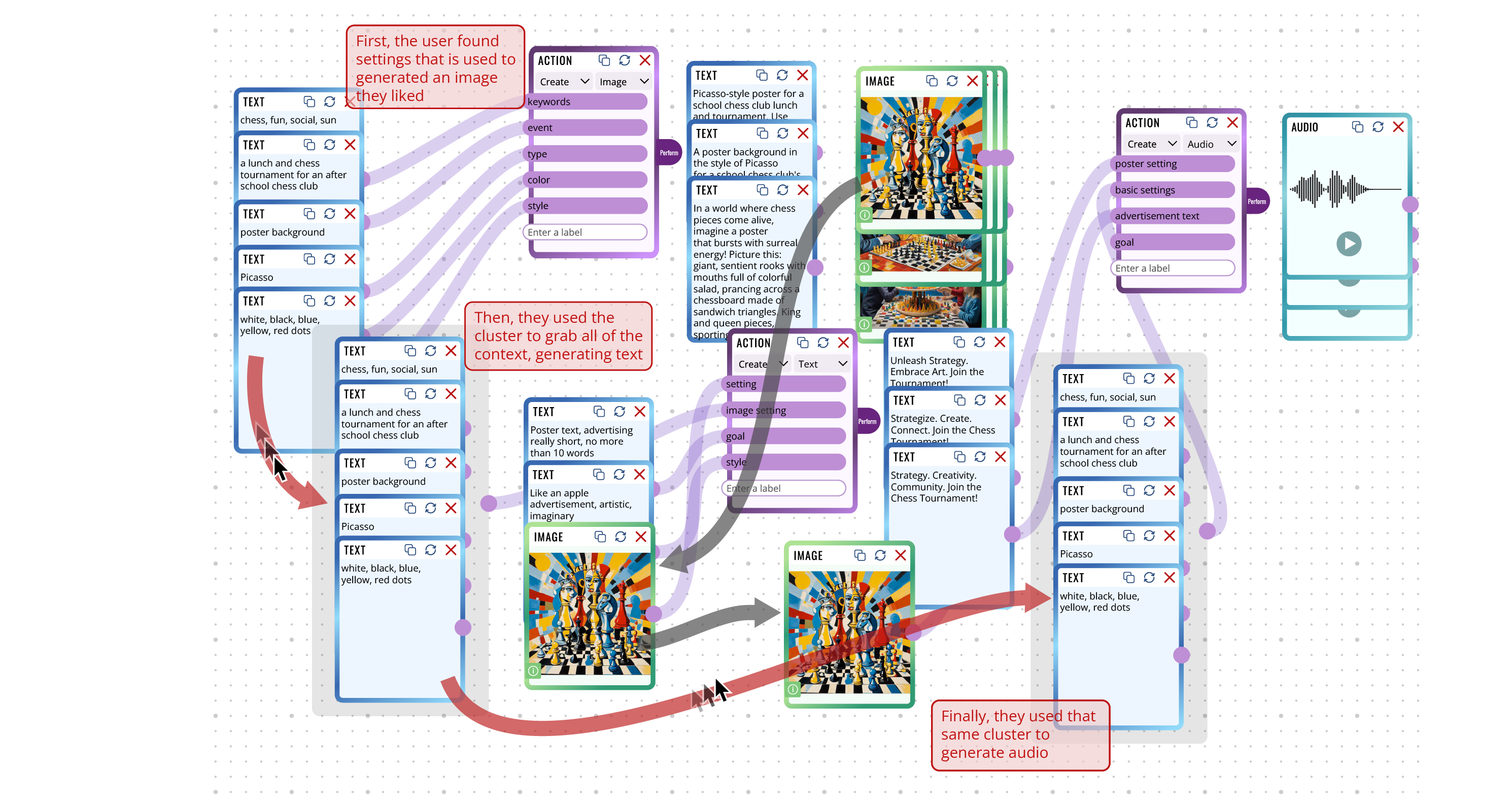}
 \caption{Some users, like \participantref{2}{6}, used \cl in creative ways, beyond the Categorization specification method.}
 \label{fig:structure_cluster}
\end{figure*}

Although originally designed to satisfy the Categorization specification method, some users (\nparticipants{1}{3},\nparticipants{2}{4}) opted to use clusters to group cards around for structural management. As seen in Figure \ref{fig:structure_cluster}, \participantref{2}{6} combined these approaches when creating a poster, text, and audio to advertise a fictional chess event; they started by finding text prompts that worked well for the poster, putting them in a cluster, and then used that cluster as input for the text and audio generation.

The single-stream conversation helped participants \paren{\nparticipants{1}{7}} keep focus on a single image: "This interface was more successful because it gave one image to work off of rather than a bunch. I wasn't working on fixing all of the images, I just had to tweak a single image" \paren{\participantref{1}{11}}.

However, participants noted difficulty in controlling this linear flow; \participantref{1}{9}, during a closed-ended task where they were asked to recreate a picture of a bird as closely as possible, stated ``[\cf] is jumping back and forth and up and down on my expectations. Sometimes, I give it a new point with some slight modification, and it will give a completely different output. I can only make a few modifications at each point... He has even lost my progress. For example, in this point I wanted to increase the number of birds from 4 to 6. Then, it increased the number from 4 to 7, and that is not correct, and then it also add a branch''.

\subsubsection{Context Management}\label{finding:context}
\begin{figure}[h]
  \centering
  \includegraphics[width=0.45\textwidth]{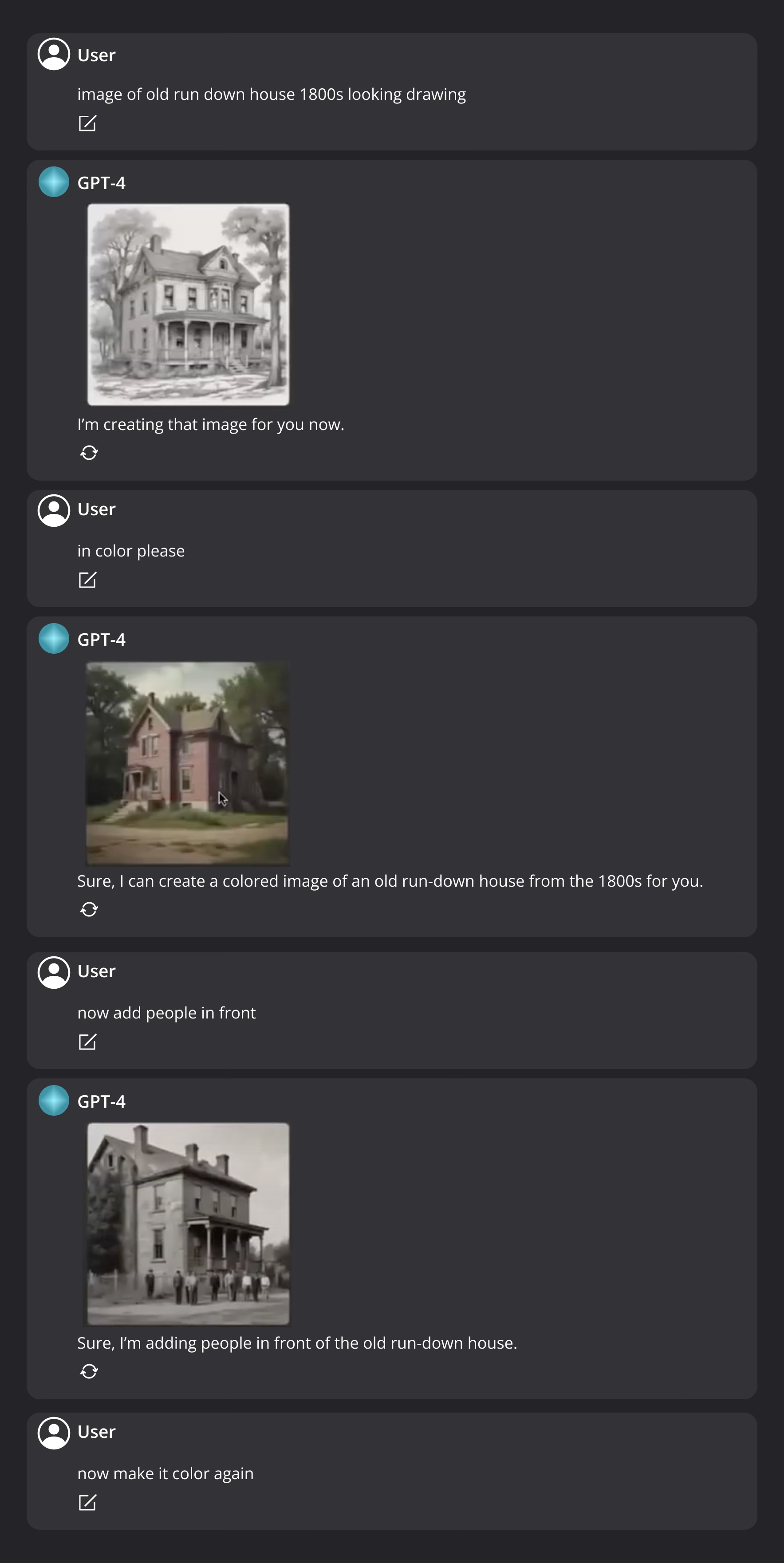}
  \caption{Users like \participantref{1}{6} had issues branching their designs in \cf.}
  \label{fig:chatflow_lose_progress}
\end{figure}

A common complaint with the baseline, \cf, was the inability to control the context of generation, including previous prompts and generated images \paren{\nparticipants{1}{6}}:

according to \participantref{1}{6} in a screen viewable at Figure \ref{fig:chatflow_lose_progress}, "\cf allows you to go step-by-step really easily, but you're also limited in that... it like takes all the the context and throws it all back in again. I'm constantly making new chats."

In \df, users found that the \ac['s] modularity made it easier to reconfigure the prompts given to the generative system \paren{\nparticipants{1}{4}}, with \participantref{1}{4} stating ``In [\cf], if I change something, the pictures would look drastically different, but the [\gc] allowed me to tweak my prompt in a more controlled way... even if I didn't love an [\ic], I know that I can add a detail to the [\ac] to get to my goal.''

\participantref{2}{6} wanted more robust history management in \df, borrowing features from the ChatBot model: ``[I would add] the feature of branching out from an input multiple times... it would be great if there was a history and backtracking stuff where you can memorize historical context---it could be optional.''

\subsubsection{Patterns Change Over Time}\label{finding:df_over_time}
A significant number of users (\nparticipants{1}{11},\nparticipants{2}{5}) expressed an evolving understanding of \df throughout the study, often articulating how their approach shifted as they became more familiar with the tool. \participantref{1}{12} reflected on this learning process:
"I guess the main reason was that I am not familiar with this tool compared to the other one, because I know how like ChatGPT generally [does] simple task step-by-steps pretty well. Now I am more familiar with [\df], if I am asked to do another task, I might just start by using a Goal Card... I find that the Goal Card is pretty effective in giving you a good starting point of everything. It gives you breakdown of the prompt, so you can iterate more easily."
Other participants also noted how their usage patterns evolved:
\begin{itemize}
\item \participantref{1}{6}: ``The cluster was helpful---didn't use much, but I would use more if I had more time. I need to be able to try it out multiple times.''
\item \participantref{1}{3}: ``The previous experienced helped me a lot compartmentalizing.''
\item \participantref{1}{1}: ``I would really like this tool if I had more practice with it---If you gave me a week free trial, I would fall in love with it.''
\end{itemize}

\subsection{RQ2: How do users approach the Specification Decomposition problem in the interfaces?}\label{finding:specification_decomposition}

\subsubsection{Specification in \cf}\label{finding:cf_specification}
In the \cf interface, all users but \participantref{1}{2} specified their requirements using a linear conversation. These messages were generally anthropomorphic, especially as users got more frustrated with \cf's ability to follow instructions, such as these from \participantref{1}{9} in Close-Ended task A: ``more closer and one branch!!!!'', ``I only see four birds?'', and ``I want one branch, why you give two branches again? also the birds are white-belle not yellow and red''.

Some users \paren{\nparticipants{1}{8}} began a task with a high-detail prompt, seeking to immediately create a potential final output. Some \paren{\nparticipants{1}{10}}, like \participantref{1}{6}, began with a basic image, adding required details one after another. Some users \paren{\nparticipants{1}{3}} downloaded some collection of favorable images, reloaded the page to reset the conversation, and uploaded the images to apply their context from a new perspective.

\participantref{1}{2} used a unique technique in which they spent over 5 minutes crafting a detailed prompt in \cf each time before generating their first output. After they received the output, they edited their original prompt, at no point taking advantage of \cf's memory capabilities.

It's worth noting that while some participants \paren{\nparticipants{1}{3}} mentioned using bullet points as one of their prompting strategies used in real-world tasks, this technique was not observed in their actual use of \cf.

\subsubsection{\ac Interactions}\label{finding:labels}

\begin{figure*}[h]
  \centering
  \includegraphics[width=\textwidth]{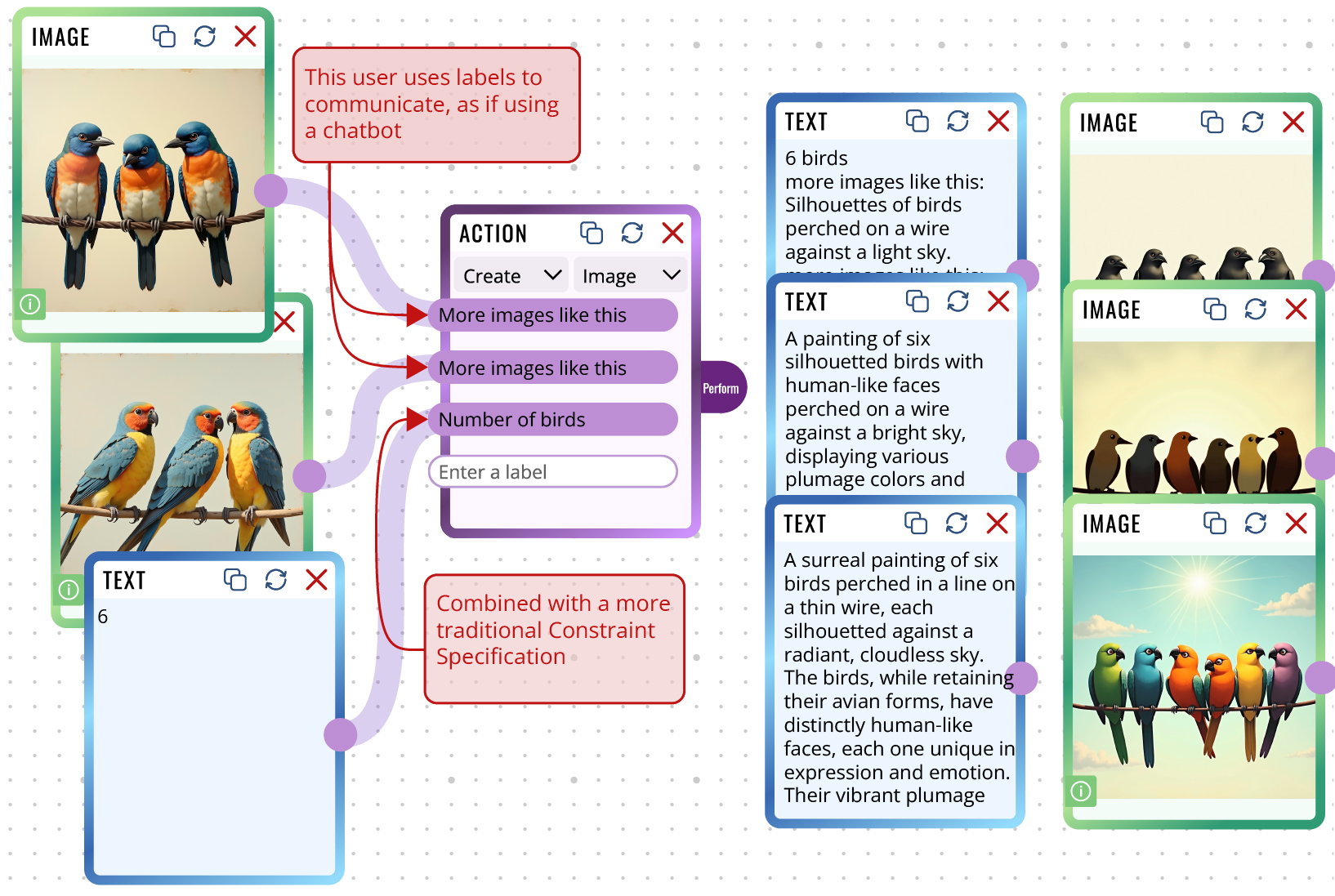}
  \caption{\participantref{1}{6} communicating the \df{} using ChatBot‑like instructions in Hard Task~B}
  \label{fig:df_chatbot}
\end{figure*}

We observed three different types of labels for input in the \ac used in each study:
\begin{enumerate}
    \item Constraint: A specific, non-interpretive condition that the output should fulfill, such as \participantref{1}{8}'s `number of birds': `6'
    \item Annotation: A label used to interpret or specify some input, such as \participantref{1}{11}'s `bird species': (image of a bird)
    \item Instruction: A natural language instruction, such as \participantref{1}{6}'s ``More images like this'' seen in Figure \ref{fig:df_chatbot}
    \item Empty: An empty label
\end{enumerate}

The `instruction' label type was not anticipated in system design, but was utilized by a few different users (\nparticipants{1}{3},\nparticipants{2}{2}). Counts of these label types used in the Multimodality study can seen in Table \ref{tab:modal}.

\begin{table*}[h]
  \centering
  \caption{Label categories and input modalities attached to an \ac{} each time a modality was generated.}
  \begin{tabular}{@{}lrrrrrrr@{}}
    \toprule
    & \multicolumn{4}{c}{\textbf{Type of Label}} & \multicolumn{3}{c}{\textbf{Type of Input}} \\
    \cmidrule(lr){2-5} \cmidrule(l){6-8}
    Modality & Annotation & Constraint & Instruction & Empty & Text & Image & Audio \\
    \midrule
    Text   & 13 & 7 & 1 & 7 & 25 & 3 & 0 \\
    Image  & 47 & 6 & 0 & 11 & 61 & 2 & 1 \\
    Audio  & 17 & 3 & 1 & 6 & 20 & 6 & 1 \\
    \bottomrule
  \end{tabular}
  \label{tab:modal}
\end{table*}

Some users (\nparticipants{1}{5},\nparticipants{2}{3}) expressed difficulty in writing labels, as verbalized by \participantref{1}{11}:
``I had trouble coming up with my own annotations for the action cards. I wasn't sure what it could take, and how specific I could get with it... It was easier to use the goal card in the beginning when I didn't quite know what I wanted.''

\subsubsection{Clusters}\label{finding:clusters}

Use of the Cluster led to discovery of concepts previously unknown by users in Close-Ended Tasks \paren{\nparticipants{1}{4}}.

In \participantref{1}{8}'s case, the Cluster even correctly identified the correct bird breed, the Zebra Finch, from Close-Ended Task A from only images previously generated in \df, without being prompted by the user. \participantref{1}{11} verbalized ``The Cluster helped it feel less overwhelming, more like I was like getting to a point rather than diverging away from it."

Some users expressed that they did not remember to use a cluster (\nparticipants{1}{2},\nparticipants{2}{3}), or avoided its use because they did not understand it \paren{\nparticipants{1}{2}}. 

Other users, however, did not find the cluster to be as useful as input; \participantref{2}{4} verbalized "I don't really understand how the clustering feature acts differently than if I were to use separate text prompts and attach them to different labels. In my head, if I'm not sure how that would change results." \participantref{2}{4} later stated ``Clustering works well to visually group the elements---the visual part of the tool is the most appealing. It does help to combine groups of prompts that you want associated with each other.''

\subsubsection{\gc Interactions}\label{finding:goal_cards}

Goal Cards was popular among participants, primarily utilized to break down high-level prompts into more manageable components. \participantref{1}{11} articulated this benefit, stating, ''The goal card helped a lot with breaking down large prompts into specific parameters, and I have the opportunity to choose which ones I want to implement.''

Some users employed Goal Cards as a task decomposition strategy, giving the system context for the entire task, not just the subtask stage. For instance, \participantref{1}{11} used a Goal Card to pose the question, ``Show me a place people want to live in''. Similarly, \participantref{2}{7} utilized a Goal Card with the label ``A child story with a twisted storyline, explained with three images and text description'' for the Children's Story task.

During the comparative study, many users \paren{\nparticipants{1}{4}} expressed a desire to provide multimodal input to the \gc. However, after this feature was implemented in the subsequent study, participants consistently used only text as input when creating an \ac.

\subsection{RQ3: How do users approach the Generative Space Exploration problem in the interfaces?}

\subsubsection{Similar performance in closed-ended tasks}\label{finding:similar_closed}

\begin{figure*}[h]
  \centering
  \includegraphics[width=\textwidth]{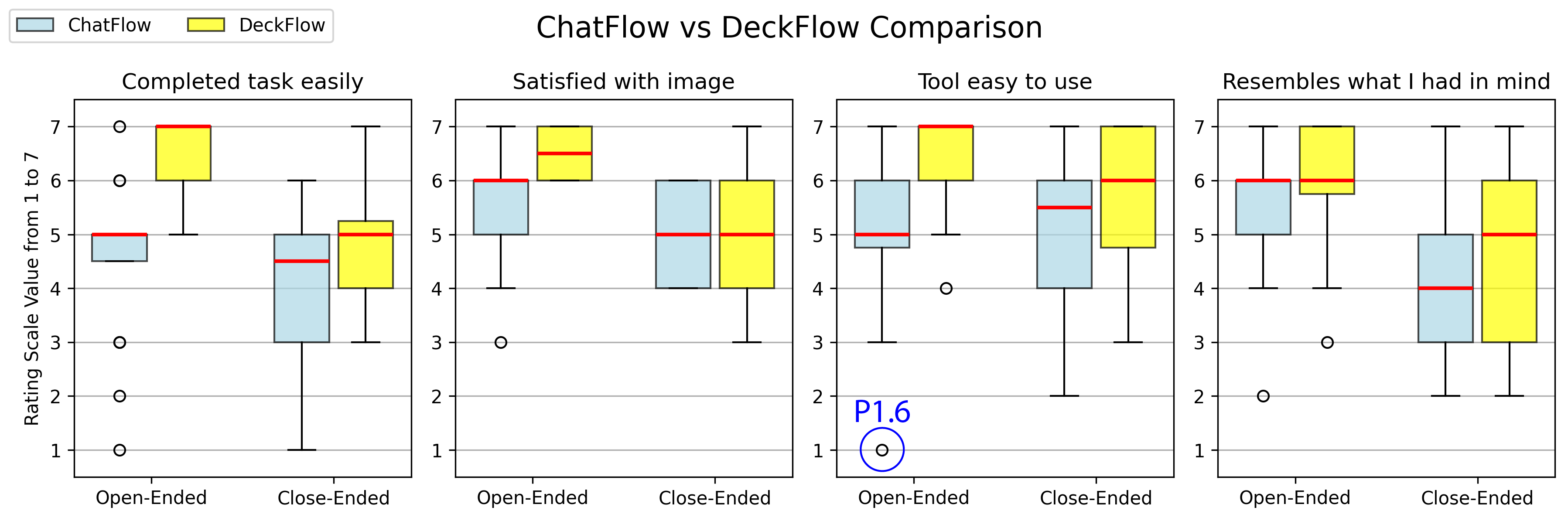}
  \caption{Rating scale results for post‑task analysis in the Comparative Study (16 open-ended, 16 close-ended). \participantref{1}{6} rated \cf{} very poorly in ease of use for the open‑ended task.}
  \label{fig:compare_chart}
\end{figure*}

In the rating scale results from the Comparative Study (Figure \ref{fig:compare_chart}), there was no clear favorite for close-ended tasks; \participantref{1}{11} stated that ``I prefer ChatFlow for [closed-ended tasks] because it's more suited for sentence prompts'', but \participantref{1}{9} disagreed, saying ``In the hard task, I definitely prefer DeckFlow because it gives me several options, and gives me an outline of what GPT expects''. In the survey, \df was more closely resembled the user's target image, but were evenly satisfied with the final image.

\begin{figure*}[h]
  \centering
  \includegraphics[width=\textwidth]{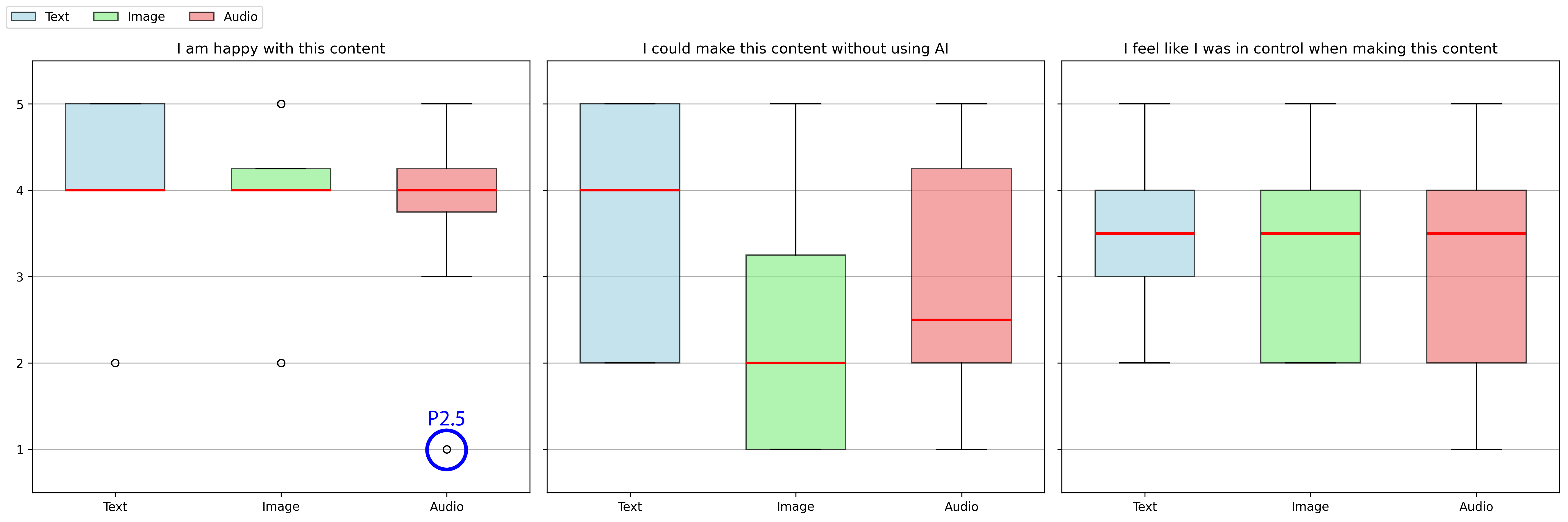}
  \caption{User evaluations of their outputs, (15 total) during the post‑task interview. \participantref{2}{5} was unhappy with the generated audio content.}
  \label{fig:retro_compare}
\end{figure*}

\begin{figure*}[h]
  \centering
  \includegraphics[width=\textwidth]{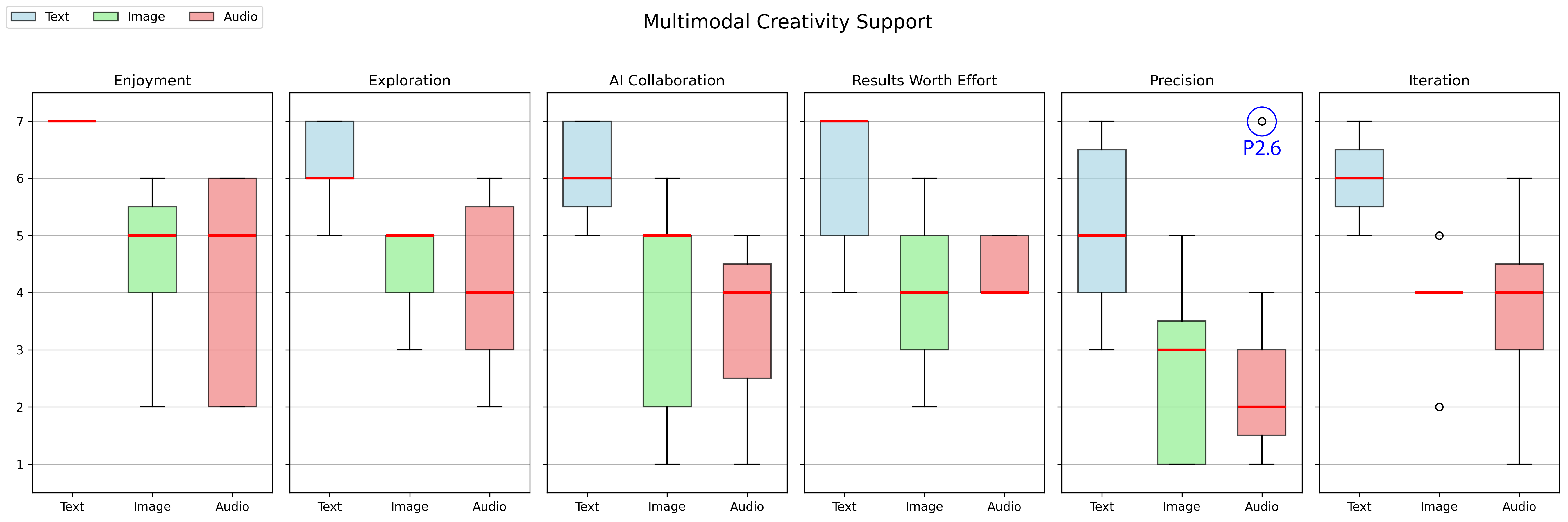}
  \caption{Adapted Creativity Support Index across modalities in the Multimodality Study \nparticipants{2}{7}.}
  \label{fig:csi_chart}
\end{figure*}

\subsubsection{\df was universally preferred in open-ended tasks}\label{finding:df_open}
As seen in Figure \ref{fig:compare_chart}, participants found \df preferable in both outcome and usability in open-ended tasks, but similar in outcome. Notably, \participantref{1}{6} gave \cf a 1 out of 7 when asked about ease of use for their open-ended task, stating ``[\cf] made more sense if I wanted to do one specific change to an image; something creative would be way easier in \df---you had such a wide array of images you got back so fast, and the prompt generation was really phenomenal. In \cf, you had to type out the prompt, or ask the model to type it out''.

\subsubsection{The role of divergence}\label{finding:divergence}
The ability of an interface to diverge from past output was noted by many users (\nparticipants{1}{8},\nparticipants{2}{4}).
\participantref{2}{7} indicated that \df's design supports this: ``I think at least the prompt part---generating image or audio---the prompts are there. I don't need to write them on my own. The creation process is a very slow process if I do it by myself. For DeckFlow, I see more possibilities more quickly. \emph{Although the image doesn't live up to my expectation, it helps me know what I want, because I know what I don't want}. Overall, I think it's great for my creative process.'' Because each of the generated rows utilize different prompting methods, some users \paren{\nparticipants{1}{5}} picked up on these differences, especially the less adherent but more creative third row: ``For the third row---it's a higher quality picture, but it doesn't really reflect the prompt ``Also, there's one thing that I have noticed is that since there are 3 roles every time, and I barely find any image from the last row relevant to what I want, although it is like a higher quality picture'' \paren{\participantref{1}{12}}.

Divergence was not always helpful, as stated by \participantref{1}{11} during their Close-Ended task: ``There was more variety in these images, and I think that worked to the detriment to me trying to do the task. I was looking for something very specific, and it didn't really quite hold.''

\subsection{RQ4: How are different modalities treated as input and output?}\label{rq:modalities}

\subsubsection{Text dominates as specification input}\label{finding:text_dominates}
Users preferred using text to specify their intent, using it for 89.5\% of their \ac inputs, and rating text much higher than images or audio for each of the modified Creativity Support Index questions \cite{creativity_support_index}: ``Sometimes, I'm not sure to what extent do the audio or images affect the output, but I can always resort to text as the input, and it's very clear to me'' \paren{\participantref{2}{3}}.

Some users \paren{\nparticipants{2}{3}} mentioned that it felt like ``Google Translate'' or the game ``telephone'' due to the intermediate context always being revealed through text and the model's gaps in modal understanding: ``I think they all go through text; it reminds me of the old Google Translate, how if you translate to different languages, especially obscure ones---Haitian Creole for example---it would go through English, and you would see the English words that they couldn't translate. \emph{The image and audio cannot directly translate to each other, but they get described by text first, which is more efficient since you don't need a separate model to communicate between that audio and the image}. At least `telephone game' results were all pretty. It was positive for entertainment since you get slightly different things, but negative for trying to make something precise'' \paren{\participantref{2}{1}}.

One notable exception is \participantref{2}{6}, who gave Audio as Input a 7 out of 7 when asked about its Precision, stating ``It's always a problem to get everything in my mind to the model. No matter with text, image, or audio---this is also a problem when I try to convey my thoughts to a real person... I can never express myself entirely---being able to express myself in different ways helped me. Like to say, if someone doesn't know Picasso, I can at least show them an image.''

\subsubsection{Text was the least interrogated output}\label{finding:text_least_interrogated}
Despite being the dominant choice for input, text only accounted for 16.8\% of the output generated. Additionally, of the \textit{final text outputs} in the Multimodality Study, 6/12 ended in incomplete sentences due to the model reaching its token limit, such as \participantref{2}{5}'s chosen text output for the Creature task: ``..., and small invertebrates. The creature often [sic]''.

\subsubsection{Most generations were images}\label{finding:images_critiqued}
In the Multimodality Study, image generation accounted for 61.0\% of all generations, consistently among each task. Participants began most tasks (8/15) with image generation, such as \participantref{2}{6}, whose creative use of the \cl was discussed in \ref{finding:cluster_structure}.

\subsubsection{Strong reactions to Audio}\label{finding:strong_audio}
Despite only averaging 2 bulk generations per task, some participants \paren{\nparticipants{2}{3}} had strong reactions to generated audio. For example, as seen in Figure \ref{fig:retro_compare}, \participantref{2}{5} was unhappy with their generated audio content which sounded vaguely like giggling and crackles: ``That was nightmarish... I'm not gonna hold it against the model. I think if it were a better model, this would be closer to what I want.'' \participantref{2}{7}, when generating audio for the Creature task, indicated they were somewhat fearful of listening to the audio, worried that it would be ``creepy'' due to the image used as input. However, after one generation, they stated ``\emph{I think the sound is what makes a creature imaginative most.} The visual part, although it may look like a different creature than what exists on earth, maybe I can imagine it from the descriptions, but the audio, I totally cannot expect. In daily life, we learn a lot from what we see, but not a lot from what we hear. I think audio is the most important part to make this creature like an alien. The audio makes me feel like the creature exists.''

\FloatBarrier

\section{Discussion}\label{sec:discussion}

\subsection{Supporting Task Decomposition Through Interactivity}
DeckFlow's infinite canvas addresses the task decomposition problem by enabling users to organize work according to their preferences. This flexibility fostered diverse usage patterns (\ref{finding:distinct}), from top-down sequential to divide-and-conquer approaches, while encouraging exploration (\ref{finding:divergence}). Components like Clusters were adapted beyond their intended purpose for canvas organization (\ref{finding:cluster_structure}), highlighting users' need for structural management. Future interfaces could enhance task decomposition through attention heatmaps from PromptCharm~\cite{wang2024promptcharm}, or spatial dimensions for parameter and prompt exploration from Automatic1111~\cite{automatic} and Dreamsheets~\cite{dreamsheets}.

Widespread AI tools, however, lack many of the features required to support these interactions. ChatGPT~\cite{IntroducingChatGPT} and Cursor\footnote{\href{https://www.cursor.com/en}{Cursor. https://www.cursor.com/en}}, for example, lack robust and glanceable incremental version control and usage flexibility. As features like the Model Context Protocol\footnote{\href{https://modelcontextprotocol.io/introduction}{Model Context Protocol. https://modelcontextprotocol.io/introduction}} become more widely integrated and new modalities enable more complicated Human-AI co-creation, it is vital that interfaces that support flexibility and glanceability develop alongside these technical accomplishments.

\subsection{Conversational Expectations}
Though DeckFlow addressed specification decomposition through structured components, users frequently approached it conversationally, potentially revealing ingrained mental models from interfaces like ChatGPT. Users  created instruction-style labels in Action Cards (\ref{finding:labels}) rather than using the intended annotation types, and used Goal Cards as if they were beginning a conversation (\ref{finding:goal_cards}). Combined with the frustration of \df lacking a centralized memory model (\ref{finding:context}), it seems that user expectations of natural language have begun to include features found in Chat interfaces. Future interface designers must either accommodate these conversational patterns, or provide clearer scaffolding for alternative interaction models.

\subsection{Multimodal Inputs and Generative Space Exploration}
DeckFlow's approach to generative space exploration through multimodal inputs showed mixed results. While supporting iterative refinement (\ref{finding:specification_decomposition}), it sometimes failed to maintain consistency in direct iteration (\ref{finding:similar_closed}), and created disconnection between modalities (\ref{finding:text_dominates}). Users strongly preferred text for specification (89.5\% of inputs) but spent most time with images (61\% of generations), and showed strongest emotional responses to audio (\ref{finding:strong_audio}).

These findings suggests that modalities can serve different roles in generation: text for precise specification, images for quickly understood output, and audio for emotional engagement. Future work could enhance exploration by giving direct control, such as attention masking~\cite{wang2024promptcharm} or model adaptation~\cite{gal2022image,hu2021lora}.

\subsection{Threats to Validity}
Our lab studies had several limitations: short tasks (10--15 min), a homogeneous participant pool (18--25 year-old CS/EE students), unfamiliar equipment, awareness of the researchers' roles, and asymmetric editing capabilities across modalities. The ChatGPT-like baseline represents just one possible comparison point in a rapidly evolving landscape. These factors limit generalizability to other tools, populations, and real-world creative workflows.

\section{Conclusion}\label{sec:conclusion}

We have presented \df, a novel interface for iterative human-AI co-design in generative content creation. \df addresses the \textbf{task decomposition problem} through an infinite canvas for parallel subtasks, the \textbf{specification decomposition problem} via Goal Cards that break into labeled Action Cards supporting multimodal inputs, and the \textbf{generative space exploration problem} by presenting structured output groups that represent different creative directions directly on the canvas.

Our evaluations demonstrated \df supports diverse workflows and improves outcomes for open-ended creative tasks. Users developed distinct patterns, had a strong preference for text-based specification despite multimodal capabilities. While our conversational baseline and \df performed similarly for closed-ended tasks, participants preferred \df for exploratory creation. The study revealed unexpected behaviors, including conversational expectations and strong emotional responses to audio generation.

As generative AI evolves, interfaces like \df will be crucial in empowering users while maintaining creative control. Future work could extend the specification design space, investigate long-term impacts on creative processes, and develop tools supporting diverse needs for content creation.

\bibliographystyle{ieeetr}
\bibliography{main}

\begin{thebibliography}{10}

\bibitem{zhou2024biasgenerativeai}
M.~Zhou, V.~Abhishek, T.~Derdenger, J.~Kim, and K.~Srinivasan, ``Bias in generative ai,'' 2024.

\bibitem{BlenderFoundation}
{Blender Foundation}, ``Blender features,'' {\em blender.org}.

\bibitem{Figma}
``Figma design features,'' {\em Figma}.

\bibitem{bragdon2010code}
A.~Bragdon, R.~Zeleznik, S.~P. Reiss, S.~Karumuri, W.~Cheung, J.~Kaplan, C.~Coleman, F.~Adeputra, and J.~J. LaViola~Jr, ``Code bubbles: a working set-based interface for code understanding and maintenance,'' in {\em Proceedings of the SIGCHI Conference on Human Factors in Computing Systems}, pp.~2503--2512, 2010.

\bibitem{whiteboard}
M.~Cherubini, G.~Venolia, R.~DeLine, and A.~J. Ko, ``Let's go to the whiteboard: how and why software developers use drawings,'' in {\em Proceedings of the SIGCHI Conference on Human Factors in Computing Systems}, CHI '07, (New York, NY, USA), p.~557–566, Association for Computing Machinery, 2007.

\bibitem{schulhoff2024promptreportsystematicsurvey}
S.~Schulhoff, M.~Ilie, N.~Balepur, K.~Kahadze, A.~Liu, C.~Si, Y.~Li, A.~Gupta, H.~Han, S.~Schulhoff, P.~S. Dulepet, S.~Vidyadhara, D.~Ki, S.~Agrawal, C.~Pham, G.~Kroiz, F.~Li, H.~Tao, A.~Srivastava, H.~D. Costa, S.~Gupta, M.~L. Rogers, I.~Goncearenco, G.~Sarli, I.~Galynker, D.~Peskoff, M.~Carpuat, J.~White, S.~Anadkat, A.~Hoyle, and P.~Resnik, ``The prompt report: A systematic survey of prompting techniques,'' 2024.

\bibitem{Promptify2023}
S.~Brade, B.~Wang, M.~Sousa, S.~Oore, and T.~Grossman, ``Promptify: Text-to-image generation through interactive prompt exploration with large language models,'' in {\em Proceedings of the 36th Annual ACM Symposium on User Interface Software and Technology}, UIST '23, (New York, NY, USA), Association for Computing Machinery, 2023.

\bibitem{comfyui}
{Comfy Org}, ``Comfyui,'' 2024.

\bibitem{arawjo2023chainforge}
I.~Arawjo, C.~Swoopes, P.~Vaithilingam, M.~Wattenberg, and E.~Glassman, ``Chainforge: A visual toolkit for prompt engineering and llm hypothesis testing,'' 2023.

\bibitem{sensescape}
S.~Suh, B.~Min, S.~Palani, and H.~Xia, ``Sensecape: Enabling multilevel exploration and sensemaking with large language models,'' in {\em Proceedings of the 36th Annual ACM Symposium on User Interface Software and Technology}, UIST ’23, p.~1–18, ACM, Oct. 2023.

\bibitem{tldrawComputer}
``Gemini {Powers} tldraw's "{Natural} {Language} {Computing}" {Experience},'' {\em Google AI for Developers}.

\bibitem{choi2023creativeconnect}
D.~Choi, S.~Hong, J.~Park, J.~J.~Y. Chung, and J.~Kim, ``Creativeconnect: Supporting reference recombination for graphic design ideation with generative ai,'' {\em arXiv preprint arXiv:2312.11949}, 2023.

\bibitem{CueFlik}
J.~Fogarty, D.~Tan, A.~Kapoor, and S.~Winder, ``Cueflik: interactive concept learning in image search,'' in {\em Proceedings of the SIGCHI Conference on Human Factors in Computing Systems}, CHI '08, (New York, NY, USA), p.~29–38, Association for Computing Machinery, 2008.

\bibitem{chung2023promptpaint}
J.~J.~Y. Chung and E.~Adar, ``Promptpaint: Steering text-to-image generation through paint medium-like interactions,'' in {\em Proceedings of the 36th Annual ACM Symposium on User Interface Software and Technology}, UIST '23, (New York, NY, USA), Association for Computing Machinery, 2023.

\bibitem{wang2024promptcharm}
Z.~Wang, Y.~Huang, D.~Song, L.~Ma, and T.~Zhang, ``Promptcharm: Text-to-image generation through multi-modal prompting and refinement,'' {\em arXiv preprint arXiv:2403.04014}, 2024.

\bibitem{suh2024luminate}
S.~Suh, M.~Chen, B.~Min, T.~J.-J. Li, and H.~Xia, ``Structured generation and exploration of design space with large language models for human-ai co-creation,'' {\em Proceedings of the SIGCHI Conference on Human Factors in Computing Systems}, 2024.

\bibitem{regaeGlassman}
T.~Zhang, L.~Lowmanstone, X.~Wang, and E.~L. Glassman, ``Interactive program synthesis by augmented examples,'' in {\em Proceedings of the 33rd Annual ACM Symposium on User Interface Software and Technology}, UIST '20, (New York, NY, USA), p.~627–648, Association for Computing Machinery, 2020.

\bibitem{dimensional_reasoning}
S.~MacNeil, J.~Okerlund, and C.~Latulipe, ``Dimensional reasoning and research design spaces,'' in {\em Proceedings of the 2017 ACM SIGCHI Conference on Creativity and Cognition}, C\&C '17, (New York, NY, USA), p.~367–379, Association for Computing Machinery, 2017.

\bibitem{Marks1997DesignGalleries}
J.~Marks, B.~Andalman, P.~Beardsley, W.~Freeman, S.~Gibson, J.~Hodgins, T.~Kang, B.~Mirtich, H.~Pfister, and W.~Ruml, ``Design galleries: A general approach to setting parameters for computer graphics and animation,'' in {\em SIGGRAPH}, pp.~389--400, 1997.

\bibitem{ODonovan2015DesignScape}
P.~O’Donovan, A.~Agarwala, and A.~Hertzmann, ``Designscape: Design with interactive layout suggestions,'' in {\em CHI}, pp.~1221--1224, 2015.

\bibitem{dreamsheets}
S.~G. Almeda, J.~Zamfirescu-Pereira, K.~W. Kim, P.~Mani~Rathnam, and B.~Hartmann, ``Prompting for discovery: Flexible sense-making for ai art-making with dreamsheets,'' in {\em Proceedings of the CHI Conference on Human Factors in Computing Systems}, CHI '24, (New York, NY, USA), Association for Computing Machinery, 2024.

\bibitem{evans2024stableaudioopen}
Z.~Evans, J.~D. Parker, C.~Carr, Z.~Zukowski, J.~Taylor, and J.~Pons, ``Stable audio open,'' 2024.

\bibitem{dang2023worldsmith}
H.~Dang, F.~Brudy, G.~Fitzmaurice, and F.~Anderson, ``Worldsmith: Iterative and expressive prompting for world building with a generative ai,'' in {\em Proceedings of the 36th Annual ACM Symposium on User Interface Software and Technology}, pp.~1--17, 2023.

\bibitem{son2023genquery}
K.~Son, D.~Choi, T.~S. Kim, Y.-H. Kim, and J.~Kim, ``Genquery: Supporting expressive visual search with generative models,'' 2023.

\bibitem{automatic}
AUTOMATIC1111, ``{Stable Diffusion Web UI},'' Aug. 2022.

\bibitem{evirgen2022ganzilla}
N.~Evirgen and X.~A. Chen, ``Ganzilla: User-driven direction discovery in generative adversarial networks,'' in {\em Proceedings of the 35th Annual ACM Symposium on User Interface Software and Technology}, UIST '22, (New York, NY, USA), Association for Computing Machinery, 2022.

\bibitem{creativity_support_index}
E.~Cherry and C.~Latulipe, ``Quantifying the creativity support of digital tools through the creativity support index,'' {\em ACM Trans. Comput.-Hum. Interact.}, vol.~21, jun 2014.

\bibitem{IntroducingChatGPT}
Nov. 2022.

\bibitem{gal2022image}
R.~Gal, Y.~Alaluf, Y.~Atzmon, O.~Patashnik, A.~H. Bermano, G.~Chechik, and D.~Cohen-Or, ``An image is worth one word: Personalizing text-to-image generation using textual inversion,'' 2022.

\bibitem{hu2021lora}
E.~J. Hu, Y.~Shen, P.~Wallis, Z.~Allen-Zhu, Y.~Li, S.~Wang, L.~Wang, and W.~Chen, ``Lora: Low-rank adaptation of large language models,'' 2021.

\end{thebibliography}

\end{document}